\documentclass[useAMS,usenatbib,usegraphicx]{mn2e}

\usepackage{amsmath}

\title[Multiple stellar populations in $z\sim1$ ETGs]{Spectral detection
of multiple stellar populations in $z\sim1$ early-type galaxies}
\author[I. Lonoce et al.]{I. Lonoce,$^{1,2}$\thanks{E-mail:
ilaria.lonoce@brera.inaf.it (OAB)} M. Longhetti,$^{1}$
P. Saracco,$^{1}$ A. Gargiulo,$^{1}$ S. Tamburri$^{1,2}$\\
$^{1}$INAF-Osservatorio Astronomico di Brera, via Brera 28, 20121 Milano, Italy\\
$^{2}$Dipartimento di Scienza e Alta Tecnologia, Universit\`a degli Studi dell'Insubria, via Valleggio 11, 22100 Como, Italy}
\begin{document}

\date{Accepted 2014 August 5.  Received 2014 July 4; in original form 2014 January 24}

\pagerange{\pageref{firstpage}--\pageref{lastpage}} \pubyear{2014}

\maketitle

\label{firstpage}

\begin{abstract}
\noindent
We present a spectroscopic analysis based on measurements of two mainly age-dependent spectrophotometric indices in the $4000$\AA $ $ rest frame region, i.e.
H+K(Ca{\small\texttt{II}}) and
$\Delta4000$, for a sample of $15$ early-type galaxies (ETGs) at $0.7<z_{spec}<1.1$, morphologically selected in the GOODS-South field.
Ages derived from the two different indices by means of the comparison with stellar population synthesis models, are not consistent with each other for at least nine galaxies
($60$ per cent of the sample), while for the remaining six galaxies, the ages derived from their global spectral energy distribution (SED) fitting are not consistent with those derived from the two indices.
We then hypothesized that the stellar content of many galaxies is made of two stellar components with different ages. The double-component analysis, performed by taking into
account both the index values and the observed SED, fully explains the observational data and improves the results of the standard one-component SED fitting in $9$ out of the $15$
objects, i.e. those for which the two indices point towards two different ages. In all of them, the bulk of the mass belongs to rather evolved stars, while a small mass fraction is many Gyr younger. In some cases, thanks to the sensitivity of the H+K(Ca{\small\texttt{II}}) index, we find
that the minor younger component reveals signs of recent star formation. The distribution of the ages of the younger stellar components appears uniformly in time and this suggests
that small amounts of star formation could be common during the evolution of high-$z$ ETGs. We argue the possibility that these new star formation episodes could be frequently
triggered by internal causes due to the presence of small gas reservoir.

\end{abstract}

\section{Introduction}

Early-type galaxies (elliptical and lenticular, hereafter ETGs) are well-known precious high-redshift candidates to trace observationally the formation and evolution of
the main structures in the Universe, since they are the most massive and so the brightest galaxies and contain most of the local observed stars and baryons
(\citealt{renzini2006} and references therein). Also the homogeneous properties of their stellar content put them in a privileged position to explore the history of their
stellar mass assembly.

Despite the lower quality and quantity of high-redshift data, direct measurements on the stellar populations of high-$z$ galaxies still remain the best solid way to constrain the
earlier evolution of ETGs \citep{vdwel, cappellari2009, guo2011, strazzullo2013}. In fact, the possibility of detecting direct pieces of evidence of stellar evolution, both through
photometric and spectroscopic data, is almost unique in a redshift range, i.e. $1<z<2$, where most of the ETG evolution is expected \citep{cimatti2004, glazebrook2004}.
A second convenience of dealing with high-$z$ data is the more reliable use of the synthetic modelling which, thanks to the relative young ages of high-$z$ sources, is
significantly less affected by degeneracy effects, e.g. age-metallicity degeneracy which is dominant in the local ETGs.

Up to now, the most used method to explore the stellar population properties of high-$z$ ETGs is the analysis of the spectral energy distribution (SED) by means of
multiband photometry \citep{daddi2005, longhetti2009, santini, sanchez2011}. However, the physical parameters extracted from the SED fitting process on synthetic models,
like in particular age, are actually only indicative mean values; thus, this method misses the detection of the possible presence of inhomogeneities in the stellar
content. In contrast, a restricted but more detailed spectroscopic analysis would be an interesting challenge to unveil that information lost in the whole SED analysis.

The important goal to be reached through the analysis of the stellar population properties of galaxies observed at $z\sim1$ in their critical state of evolution, is to
understand if all of their stars are coeval and still passively evolving, as they are usually modelled with $z_{form}>2-3$ \citep{renzini2006}, or there
are some pieces of evidence of later star forming episodes or minor merging events which led to composite stellar populations already at $z\sim1$.

The hidden complexity of the star formation history of ETGs has been revealed by many works on local ETGs \citep{coccato2010, panuzzo2011, kaviraj2012, rocca2013}, and
at intermediate redshift \citep{treu2002}, where
different kinds of signature suggest that elliptical galaxies have possibly undergone rejuvenation episodes in their recent history. Also at high redshift, with both
archaeological and direct measurements, the necessity of introducing multiple stellar populations, with different stellar properties, in the used
modelling has been found \citep{garg12, huang2013}. The main stellar properties involved in the differentiation of stellar components are age and metallicity. However, age is surely
the first stellar parameter which should be investigated on high-$z$ ETGs. In fact, direct spectroscopic measurements of stellar metallicity on high-$z$ ETGs are still a
hard issue; moreover, \citet{garg12} found that age gradients are necessary to reproduce the majority of the observed high-$z$ colour gradients.

Coexistent different stellar populations in ETGs are expected from the theoretical models on galaxy formation and evolution: minor mergers and small accretion events instead of
major mergers, in fact, are predicted to be frequent until $z=0$ \citep{naab2009, oser2010, oser2012}. This inside out formation scenario, however, predicts that for high mass galaxies,
accreted stars have been formed in the same epoch of ``\emph{in situ}'' stars \citep{oser2010}, assembling thus composite stellar populations of rather coeval stellar content, though the minor mergers have occurred at different
times.

So the attention now should be focused on finding some observables which allow one to understand how homogeneous are the stellar populations of elliptical galaxies and
how important are the mass fraction and the age of the accreted stars.
A spectroscopic analysis on a sample of high-$z$ ETGs can surely reveal precious details on their stellar properties, to shed light on the composition of the stellar
populations of galaxies in their earlier phases of evolution. Previously, other pioneer works \citep{onodera, jorgensen} have been devoted to the analysis of
spectroscopic high-$z$ data, revealing many differences in the stellar properties of $z>1$ objects with respect to local ETGs, such as the presence	of extreme values of spectral
indices not foreseen by models \citep{onodera}. In this paper we present our spectroscopic analysis of a small sample of ETGs based on some of the first high quality
optical spectroscopic data at $z\sim1$ in the GOODS-South field, which, thanks to the measurements of reliable age-dependent spectrophotometric indices in the $4000$\AA $ $
rest frame region, i.e. the H+K(Ca{\small\texttt{II}}) and $\Delta4000$, has led to interesting results on this topic.

The paper is organized as follows: in Section \ref{sample} we introduce our sample and describe the spectroscopic data with the main passes of the data reduction. In
Section \ref{indici} we present the definition and the measurements of the spectrophotometric indices which are the basis of our spectral analysis. The comparison of
the obtained results with synthetic stellar population models is widely discussed in Section \ref{analisi}, together with the description of our proposed double-component
models. In Section \ref{discussione} we discuss our results through the comparison with other works based on high-$z$ spectral analysis. We propose the summary and our
conclusion in Section \ref{conclusione}.

Throughout this paper, we assume a standard cosmology with H$_0=70$ km s$^{-1}$ Mpc$^{-1}$, $\Omega_m=0.3$ and $\Omega_{\Lambda}=0.7$. All photometric magnitudes are
expressed in the AB system.

\section{The sample}
\label{sample}
The sample consists of $15$ ETGs at $0.7<z_{spec}<1.1$ morphologically selected from a catalogue complete to $K\simeq22$  (Tamburri et al., in preparation),
in the southern field of the Great Observatories Origins Deep Survey (GOODS-South v2; Giavalisco et al $2004$).
The selection criterion of this sample was basically the availability of sufficient high-quality (S/N $>5$) spectroscopic data to perform the measurements of the main
spectrophotometric indices in the $4000$\AA $ $ break region, which is the aim of this work. Thus, starting from the complete sample of $196$ morphologically selected ETGs
with $K<22$, only $15$ could provide a high-S/N spectrum suitable for the present study (see Section \ref{spectroscopic} for details).

All the selected galaxies have the 14 bands GOODS-South survey coverage \citep{santini}: deep optical images taken from four \emph{Hubble Space Telescope}-Advance Camera for Surveys
bandpasses ($F435W$, $F606W$, $F775W$ and $F850LP$); photometric data provided by extensive observations of European Southern Observatory (ESO) telescopes both in the optical (three $U$-band filters)
and in the near-infrared ($J$, $H$ and $K$ filters), and by the four \emph{Spitzer-Infrared Array Camera bands} ($3.6$, $4.5$, $5.8$ and $8.0\mu$m).

Thanks to these multiwavelength data we were able to perform the global SED fitting of all the galaxies of the sample in order to extract
the global stellar population properties. The SED fitting process was carried out assuming the stellar population synthesis model of Bruzual and Charlot \citep{bc03}
(hereafter BC03) with a Chabrier initial mass function (IMF)\citep{chabrier}. We adopted star formation histories with five exponentially declining star formation rates with $e$-folding time $\tau=$[$0.1$, $0.3$, $0.4$, $0.6$, $1.0$] Gyr and assumed solar metallicity Z $=0.020$. Dust extinction, following \citet{calzetti}, was applied in the
range $0<A_{v}<2$ mag. In Table \ref{tab:fotom}, we report the stellar parameters obtained from the SED fitting of the 15 ETGs: photometric ages $Age_{phot}$ of the
stellar populations, dust extinction $A_{v}$, stellar masses $\mathcal{M}_*$ and star formation time-scale $\tau$.

For five objects out of the whole sample (those at the highest spectral resolution) the velocity dispersion measures are available \citep{vdwel}. Furthermore,
$7$ out of the $15$ target galaxies (those at $z> 0.9$) are included in the complete sample of the GOODS-South field ETGs at $z>0.9$ \citep{34etg}.

\begin{table*}
 \centering
 \begin{minipage}{140mm}
 \caption{Sample of $15$ ETGs. Data extracted from photometric analysis: stellar population age (A$_{phot}$), dust extinction (A$_v$), logarithm of the stellar mass
 (log($\mathcal{M}_*$)) and star formation time-scale ($\tau$). $\sigma$ is the velocity dispersion measured by \citet{vdwel}. Units of right ascension are hour,
 minutes and seconds, and unites of declination are degrees, arcminutes and arcseconds.}
 \label{tab:fotom}
 \begin{tabular}{lcccccccc}
 \hline
 ID-MUSIC & RA & Dec & $z_{spec}$ & Age$_{phot}$ & Av & log($\mathcal{M}_*$) &$\tau$& $\sigma$  \\
 & & &                        & (Gyr)&(mag)&(M$_{\odot}$)&(Gyr)&(km s$^{-1}$) \\
 \hline
 1192$^a$ & 3:32:25.16 & -27:54:50.1 & 1.089 &0.7&1.0&11.01&0.1&231$\pm$15\\
 1382$^a$ & 3:32:22.93 & -27:54:34.3 & 0.964 &1.4&0.4&11.08&0.1&200$\pm$9 \\
 1950$^a$ & 3:32:26.29 & -27:54:05.0 & 1.044 &1.9&0.3&10.75&0.3&300$\pm$30\\
 1837$^a$ & 3:32:19.29 & -27:54:06.1 & 0.964 &3.0&0.4&11.40&0.4&336$\pm$18\\
 2694$^a$ & 3:32:31.37 & -27:53:19.1 & 1.135 &2.5&0.3&11.40&0.3&232$\pm$19\\
 9066$^a$ & 3:32:33.06 & -27:48:07.5 & 1.188 &2.0&0.6&10.58&0.3&-\\
 11539$^a$& 3:32:37.19 & -27:46:08.1 & 1.096 &2.3&0.5&11.42&0.3&-\\
 10020 & 3:32:15.81 & -27:47:13.6 & 0.738 &1.8&0.5&10.85&0.3&-\\
 10960 & 3:32:19.24 & -27:46:32.2 & 0.737 &2.4&0.3&10.76&0.3&-\\
 11225 & 3:32:14.44 & -27:46:24.5 & 0.736 &1.4&0.2&10.22&0.1&-\\
 9792  & 3:32:18.01 & -27:47:18.6 & 0.734 &3.25&0.5&11.58&0.4&-\\
 13386 & 3:32:17.49 & -27:44:36.7 & 0.734 &1.7&0.3&10.31&0.3&-\\
 9838  & 3:32:17.94 & -27:47:21.5 & 0.732 &3.25&0.2&11.08&0.4&-\\
 17044 & 3:32:37.38 & -27:41:26.2 & 0.672 &3.25&0.1&11.21&0.4&-\\
 7424  & 3:32:39.54 & -27:49:28.4 & 0.669 &2.6&0.2&10.54&0.4&-\\
 \hline
\end{tabular}
 $^a$objects belonging to $34$ ETGs of \citet{34etg}.
\end{minipage}
\end{table*}

\subsection{Spectroscopic data}
\label{spectroscopic}
As already mentioned, our sample of ETGs was collected following the disponibility of high-S/N optical spectra
in the GOODS-South field for galaxies morphologically confirmed as ETGs with $z\sim1$. High-quality spectra come from three different observing campaigns:
one summarized in \citet{mignoli} (K20 survey), one in \citet{popesso} and in \citet{vdwel}. We could not use spectra coming from the campaign VLT/FORS2
spectroscopy in the GOODS-South field reported in \citet{vanz05,vanz08}, because observations have been made without the cut order filter OG590+32,
so they have not a solid continuum shape calibration necessary for the measurement in particular of the $\Delta4000$ index.

The K20 survey is a near-infrared-selected redshift survey targeting galaxies ($K_s < 20$, Vega system) in two fields, one of which has been later included in the GOODS-South area.
From this precious data base of public galaxy spectra (http://vizier.u-strasbg.fr/viz-bin/VizieR?-source=J/A$\%2$BA/$437$/$883$), we have searched for the highest S/N
spectra of $z\ge0.7$ ETGs in the 4000\AA $ $ break region rest frame. We finally extracted eight ETGs in the redshift range $0.73<z<1.2$.
Spectroscopic observations of these objects come from VLT-FORS1 in MOS mode and VLT-FORS2 in MXU mode (multi-object spectroscopy with exchangeable masks), with a
set of grisms $150I$, $200I$ and $300I$ providing dispersion of $5.5$, $3.9$ and $2.6$ \AA/pixel and spectral resolution of R $=260$, $380$ and $660$ respectively.
The observation period of the K20 spectroscopic survey was confined in $20$ nights during four years (1999-2000, 2002). For more details, see \citet{mignoli}.

Our next source of optical spectra was the second campaign of the ESO-GOODS spectroscopic programme operated with VLT-VIMOS (VIsible MultiObject Spectrograph) and
reported in \citet{popesso}. From the thousands of available spectra we have looked for the ones with longer exposure time (exp-time $>5$h) in order to reach
the highest S/N for $z>0.7$ ETGs. Moreover, our searching has been focused only on median resolution (MR - resolution R $=580$, dispersion 2.5\AA/pixel) grism data,
necessary to perform the measurements of the spectrophotometric indices. We found only $2$ suitable objects in the redshift range $z\sim0.7-1.0$.
Observations were carried on during the winter of 2004-2005 for both objects. For deeper details, see \citet{popesso}.

Finally, the last five objects of our sample were analysed by means of the optical spectra coming from observations performed by \citet{vdwel} with FORS2 in MXU mode
on VLT telescope. They used the GRIS-600z grism together with the OG590 order separation filter, with a resolution of R $\sim1390$, that led to a binned spectral
dispersion of $1.6$ \AA/pixel and a wavelength coverage from about $6500$ to $11000$\AA $ $ (that means around the spectral region of the Balmer/$4000$\AA $ $ break for
galaxies at $z\sim1$). The observations of the five objects were carried out in the period from 2002 September to 2003 March, with a mean exposure time of
about 11 h for object (for details, see \citealt{vdwel}).

All the details of the spectroscopic data are summarized in Table \ref{tab-spectradata}.

\begin{table*}
 \centering
 \begin{minipage}{140mm}
 \caption{Information about the spectroscopic data used in this work: spectroscopic instrument, grism, dispersion, resolution and the observing run.}
 \label{tab-spectradata}
 \begin{tabular}{lccccc}
 \hline
 ID-MUSIC & Instrument & Grism & Dispersion & Resolution & Observing run            \\
          &            &       &(\AA/pixel) &            &\\
 \hline
 1192     & VLT-FORS2  & 600z  & 1.6        &  1390      & \citet{vdwel}              \\
 1382     & VLT-FORS2  & 600z  & 1.6        &  1390      & \citet{vdwel}              \\
 1950     & VLT-FORS2  & 600z  & 1.6        &  1390      & \citet{vdwel}              \\
 1837     & VLT-FORS2  & 600z  & 1.6        &  1390      & \citet{vdwel}              \\
 2694     & VLT-FORS2  & 600z  & 1.6        &  1390      & \citet{vdwel}              \\
 9066     & VLT-FORS2  & 300I  & 2.6        &  660       & \citet{mignoli}            \\
 11539    & VLT-FORS2  & 200I  & 3.9        &  380       & \citet{mignoli}            \\
 10020    & VLT-FORS1  & 150I  & 5.5        &  260       & \citet{mignoli}            \\
 10960    & VLT-FORS2  & 150I  & 5.5        &  260       & \citet{mignoli}            \\
 11225    & VLT-FORS2  & 200I  & 3.9        &  380       & \citet{mignoli}            \\
 9792     & VLT-FORS1  & 150I  & 5.5        &  260       & \citet{mignoli}            \\
 13386    & VLT-FORS2  & 200I  & 3.9        &  380       & \citet{mignoli}            \\
 9838     & VLT-FORS1  & 150I  & 5.5        &  260       & \citet{mignoli}            \\
 17044    & VLT-VIMOS  & MR    & 2.5        &  580       & \citet{popesso}            \\
 7424     & VLT-VIMOS  & MR    & 2.5        &  580       & \citet{popesso}            \\
 \hline
\end{tabular}
\end{minipage}
\end{table*}

\subsection{Data reduction}
\label{datareduction}
Spectra coming from \citet{popesso} and \citet{vdwel} have been reduced starting from raw observed data. For K20-survey spectra, we used
directly the online material (from Vizier: ``K20 survey: spectroscopic catalogue \citep{mignoli}''). Standard spectral data reduction has been applied using {\small\texttt{IRAF}}
software tools.
Particular care has been devoted to the relative spectral flux calibration, which has been achieved using spectrophotometric standard stars observed in the same nights of the
targets; in particular, for VLT-FORS2 data \citep{vdwel}, we took the mean sensitivity function built with standard stars of different spectral type, which reveal small
intrinsic differences in their shape. Moreover, still with the aim of obtaining the best continuum shape calibration necessary for a solid measure of spectral indices,
we applied a further continuum calibration on the VLT-FORS2 spectra: indeed we noted a systematic distortion of the spectrum shape for objects whose slits are
shifted with respect to the central position, along the dispersion direction (see details in Longhetti et al., in preparation).
The amount of the flux distortion is proportional to the distance of the slit from the central position and, in the particular case
of our VLT-FORS2 objects, the maximum offset slit position leads to a variation up to about $30$ per cent of the flux.
We corrected for this further distortion using flat-field lamp spectra as standard spectral sources, available for all the slits. The ratio between the wavelength-calibrated flat-field
spectrum obtained in a central position and the one obtained through each slit in offset position has been used for this further correction.
The final check of the obtained continuum shape accuracy in the spectral region involved in the measure of the chosen spectrophotometric indices, was to overlap the
final 1D spectra with the photometric points in the optical bands. The agreement is reached within 1-$\sigma$ error from the photometric points.

\begin{figure*}
\includegraphics[width=5.83cm]{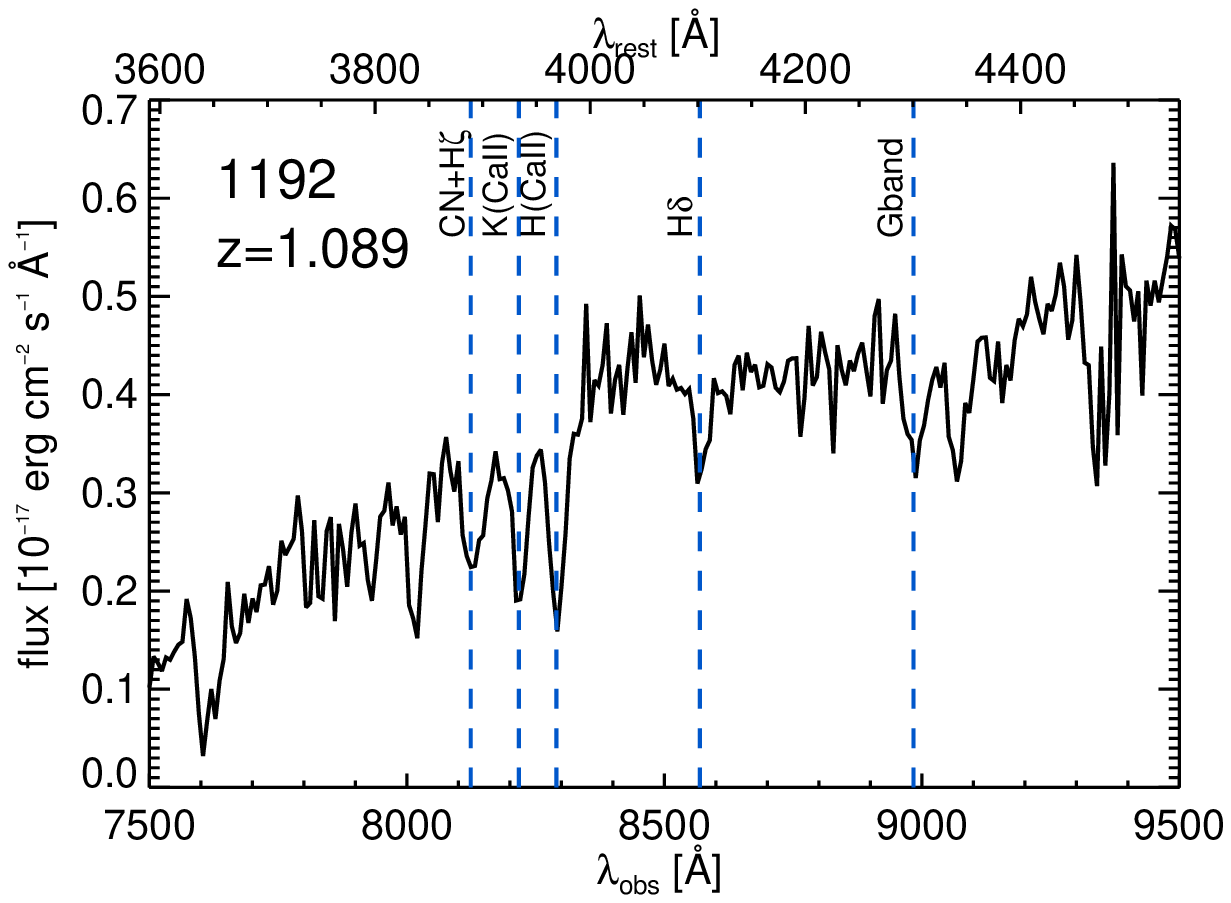}
\includegraphics[width=5.83cm]{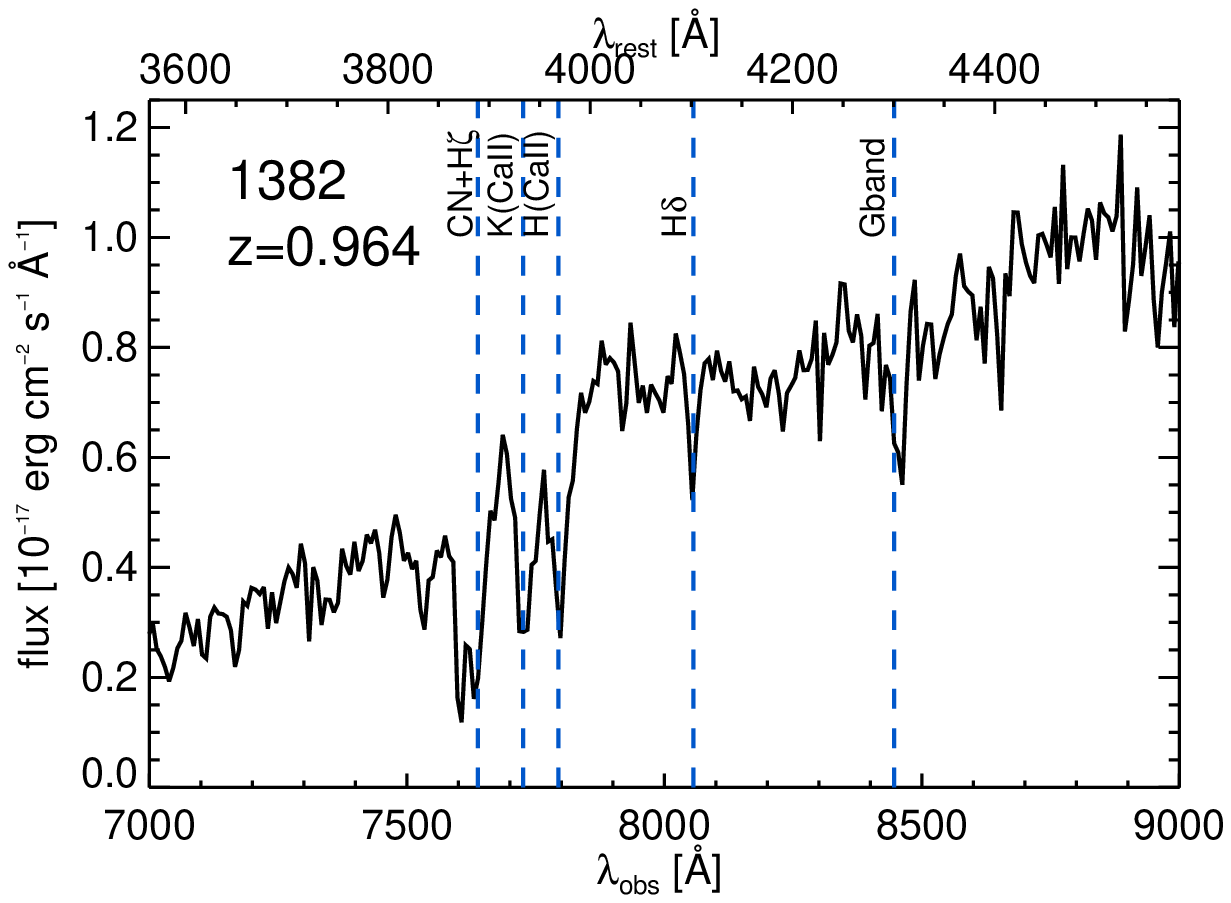}
\includegraphics[width=5.83cm]{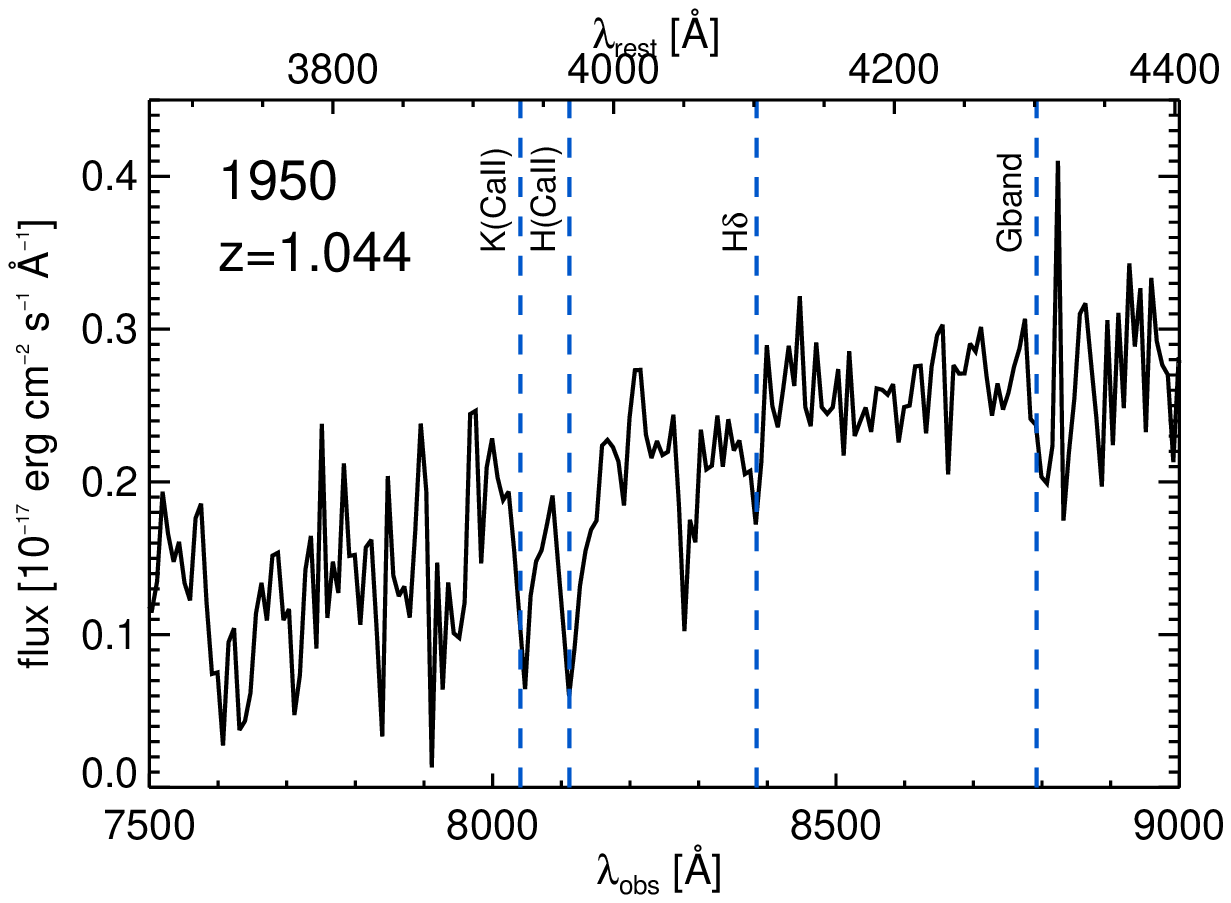}
\includegraphics[width=5.83cm]{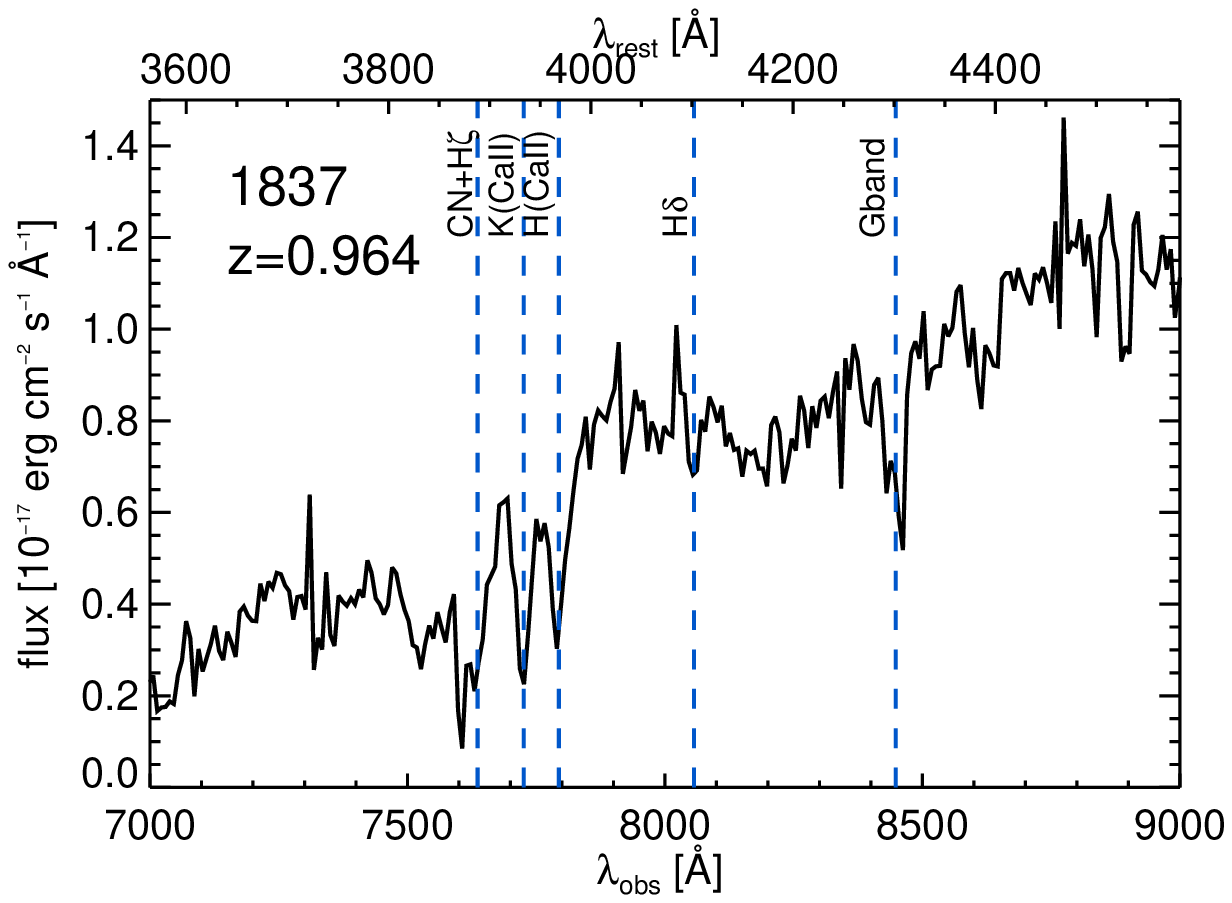}
\includegraphics[width=5.83cm]{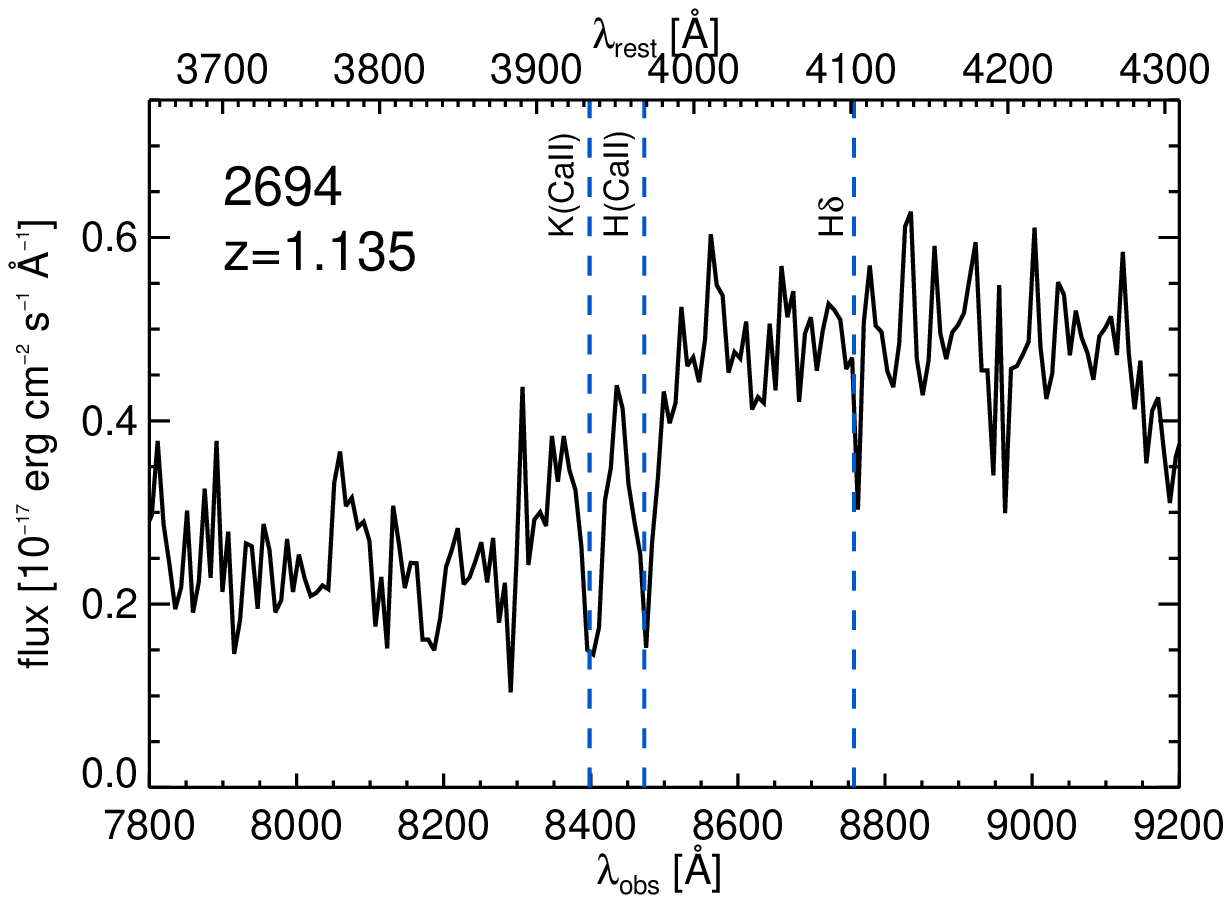}
\includegraphics[width=5.83cm]{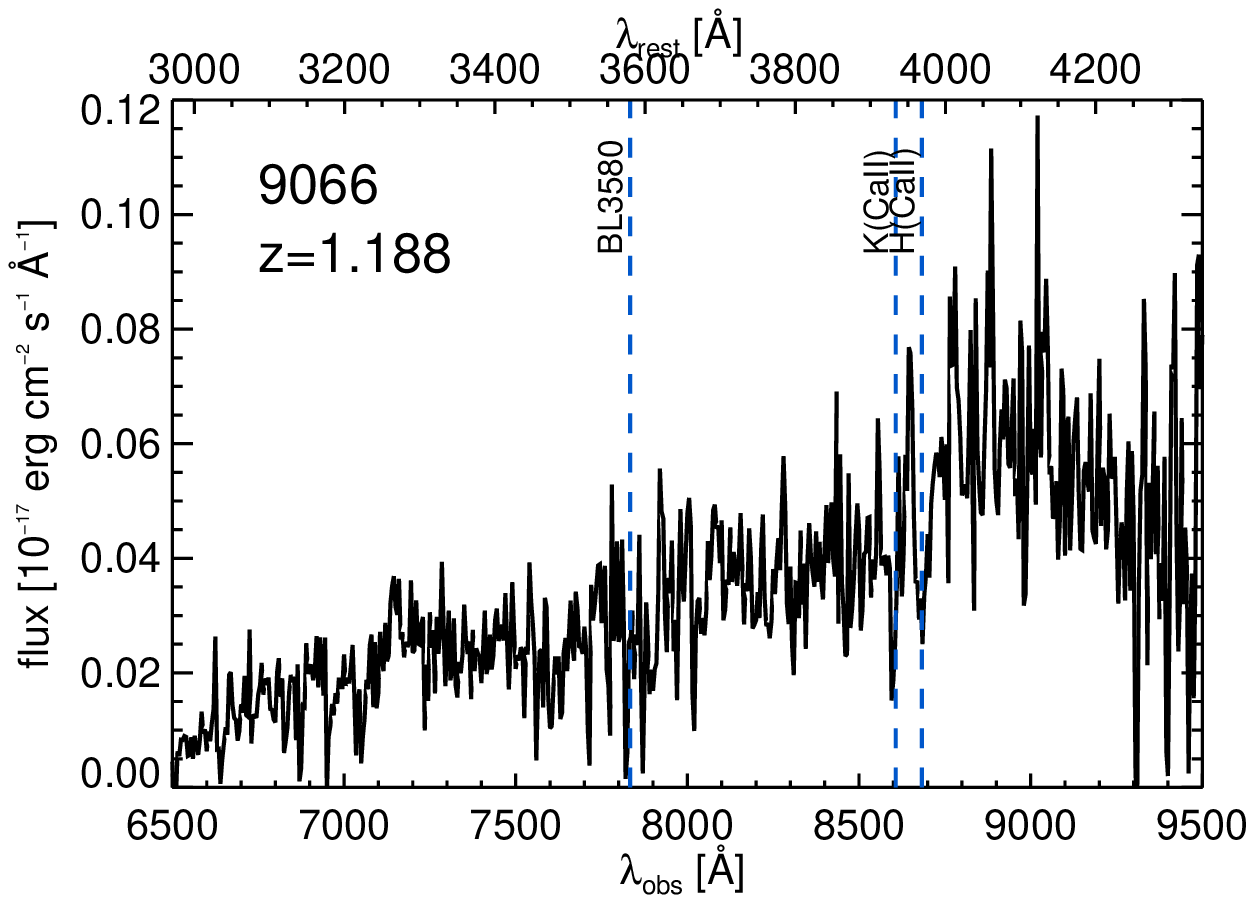}
\includegraphics[width=5.83cm]{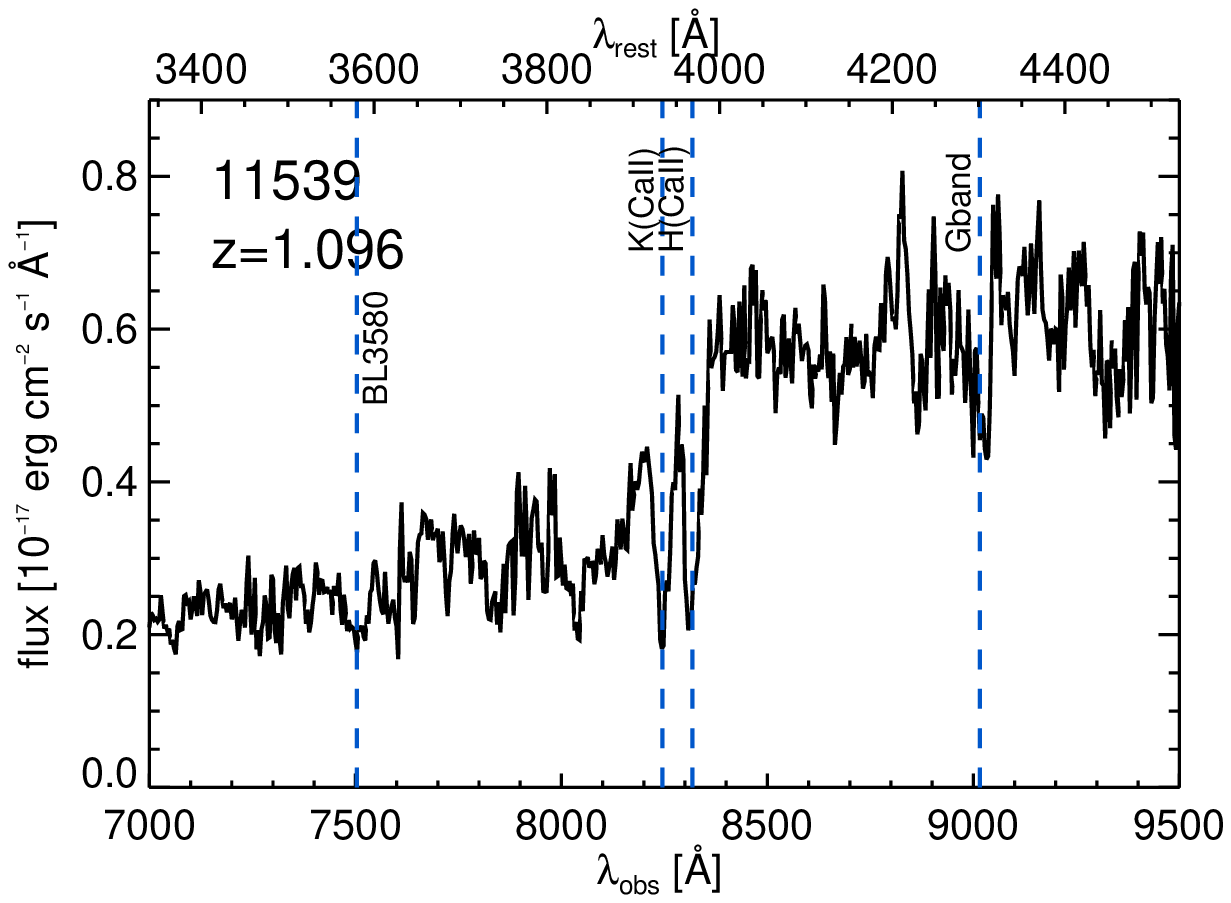}
\includegraphics[width=5.83cm]{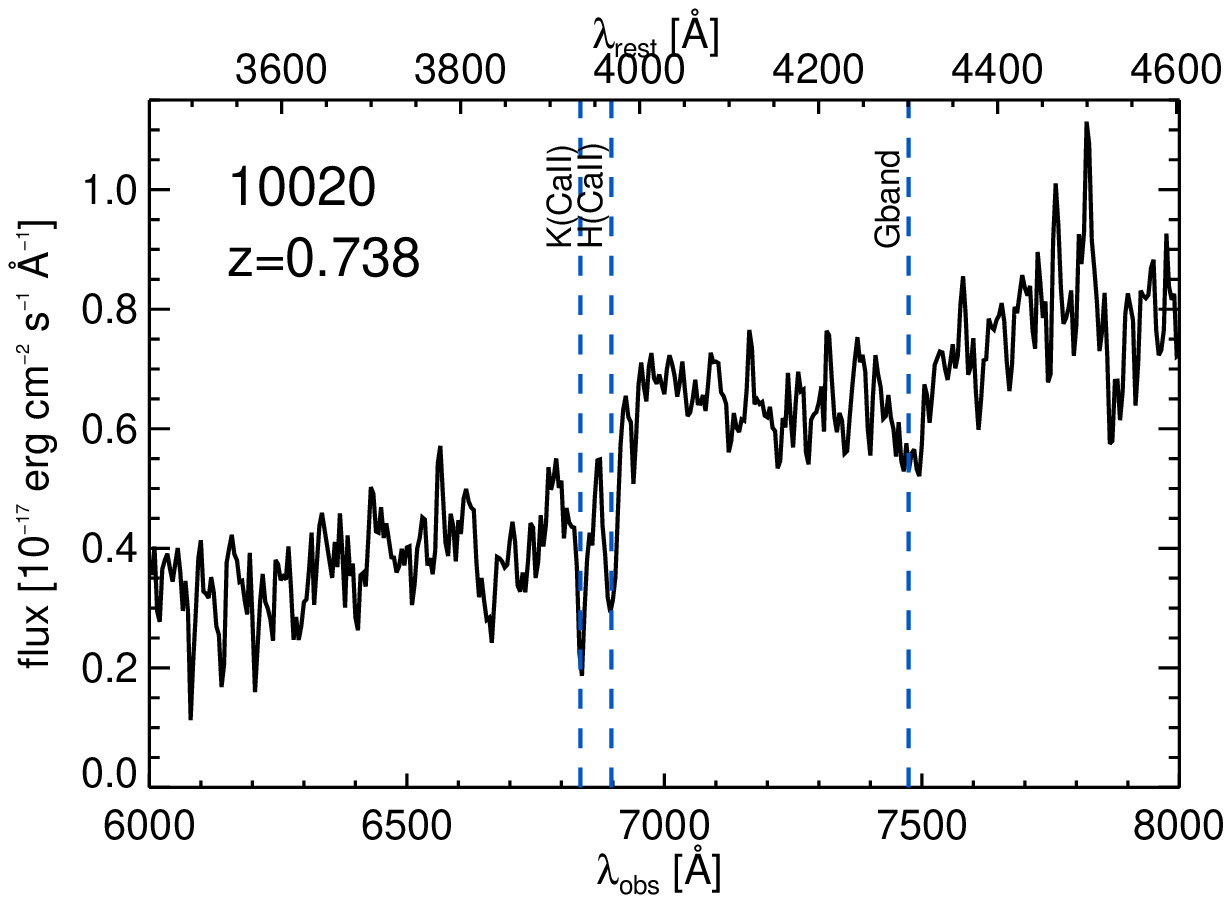}
\includegraphics[width=5.83cm]{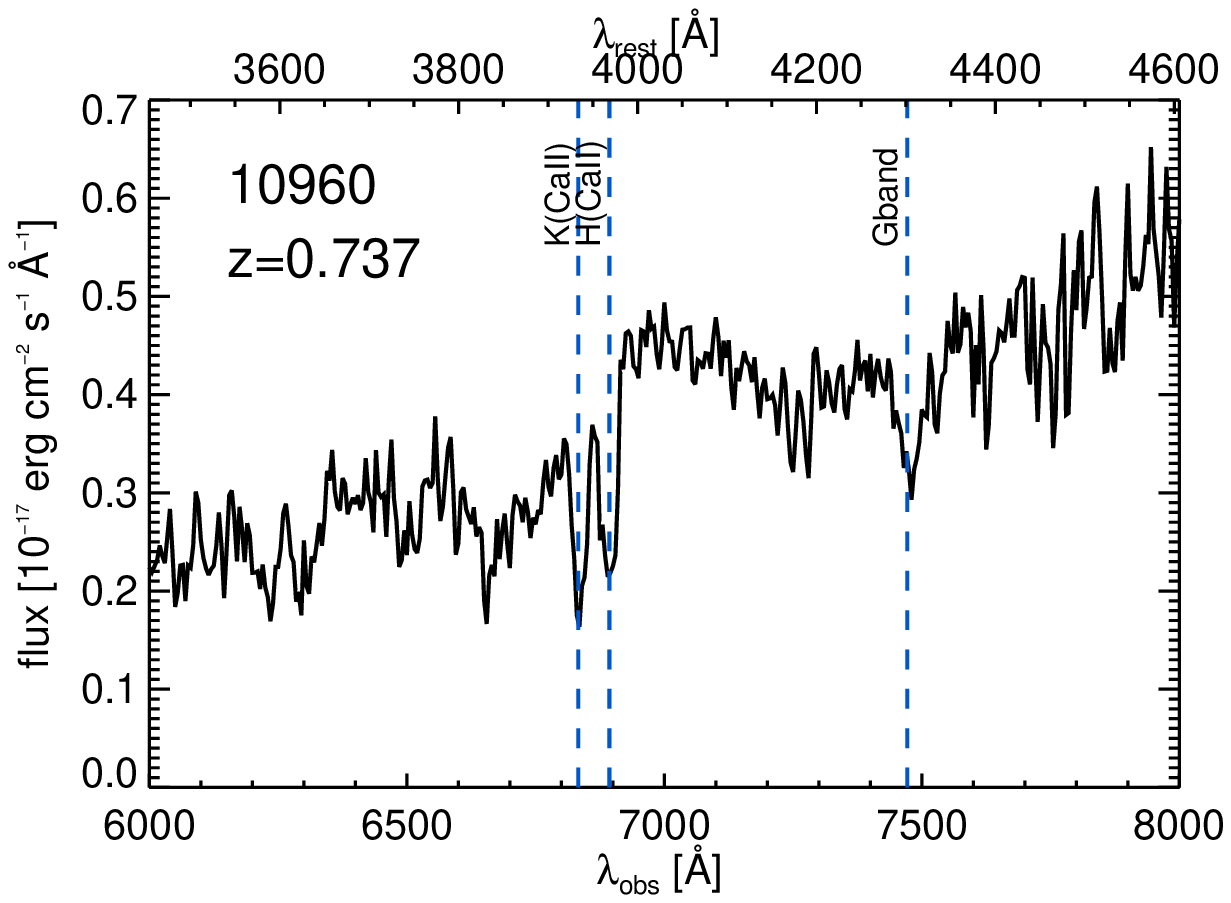}
\includegraphics[width=5.83cm]{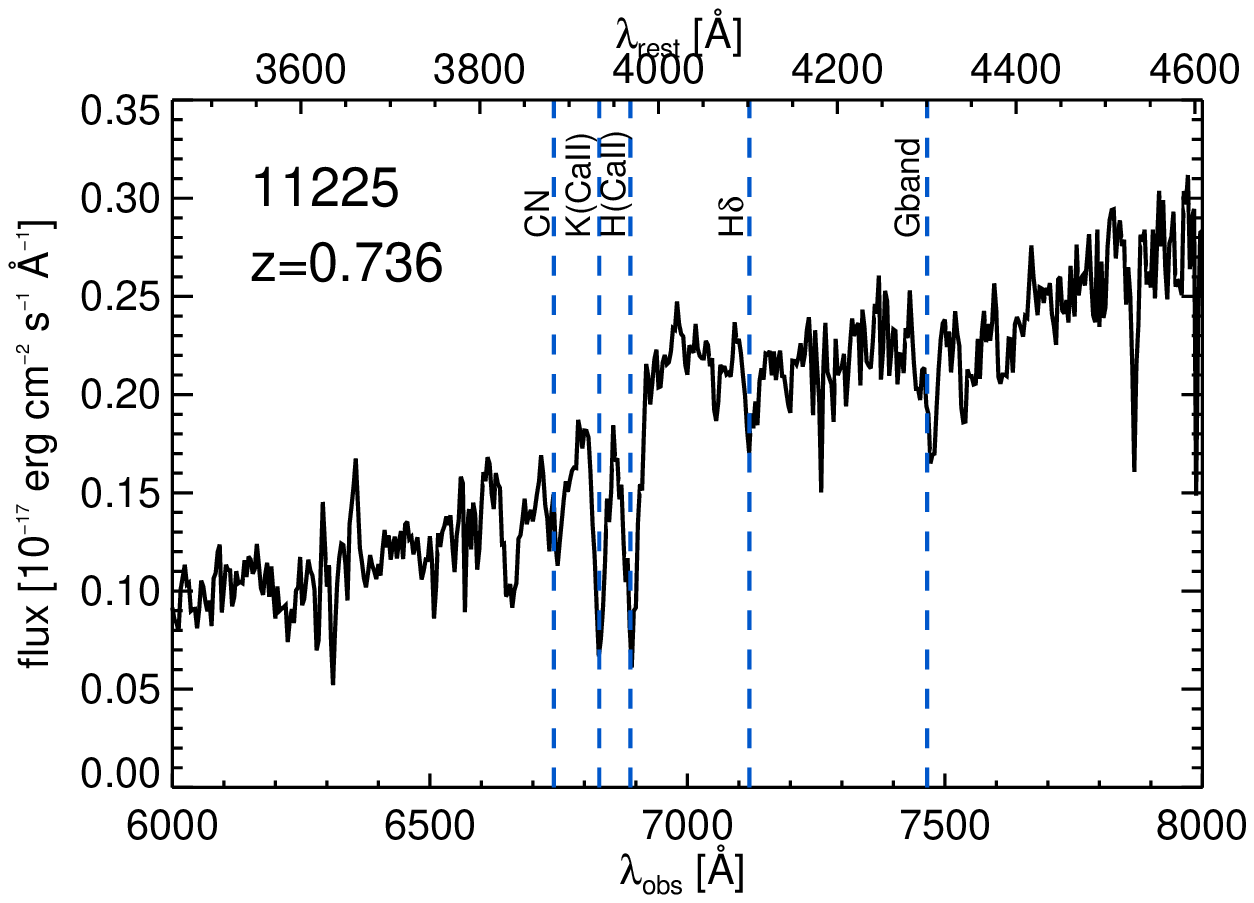}
\includegraphics[width=5.83cm]{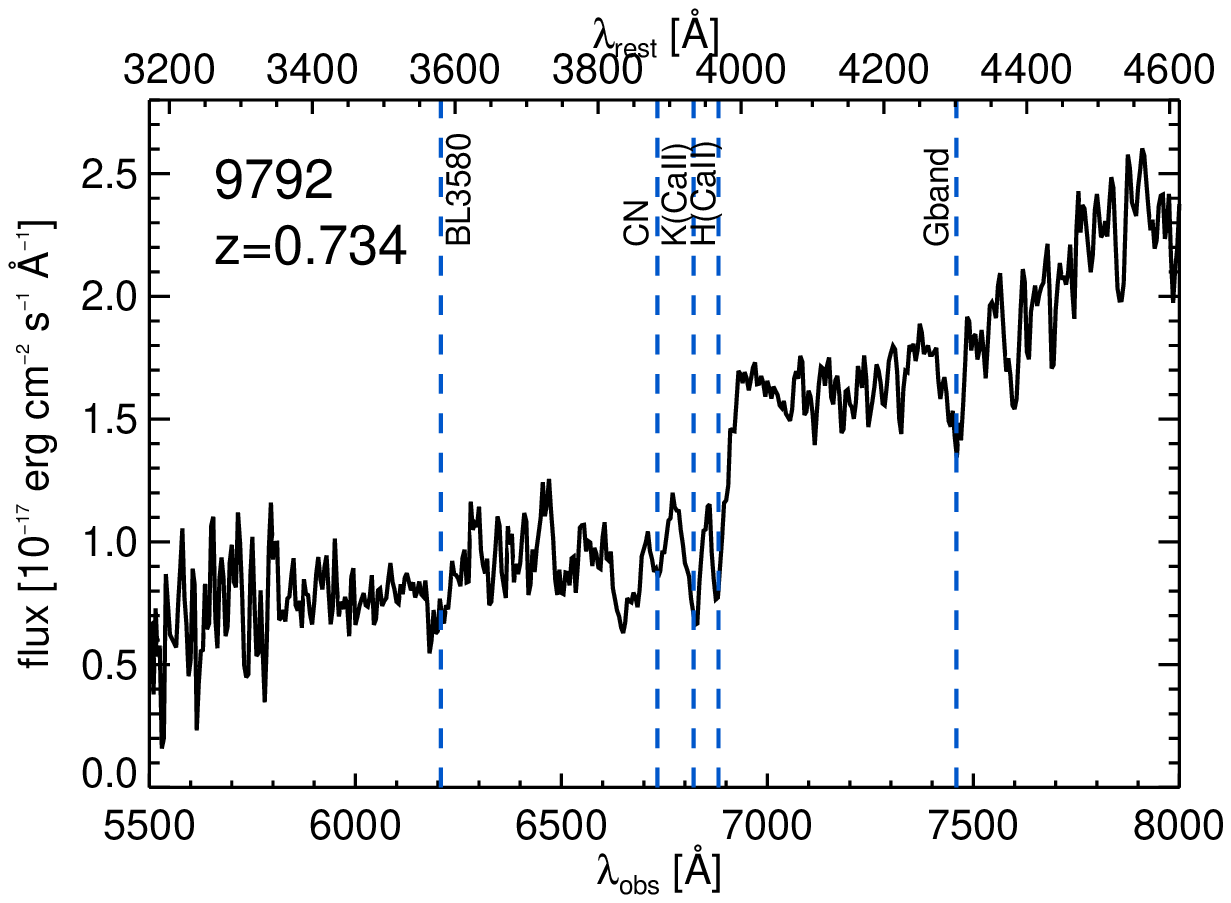}
\includegraphics[width=5.83cm]{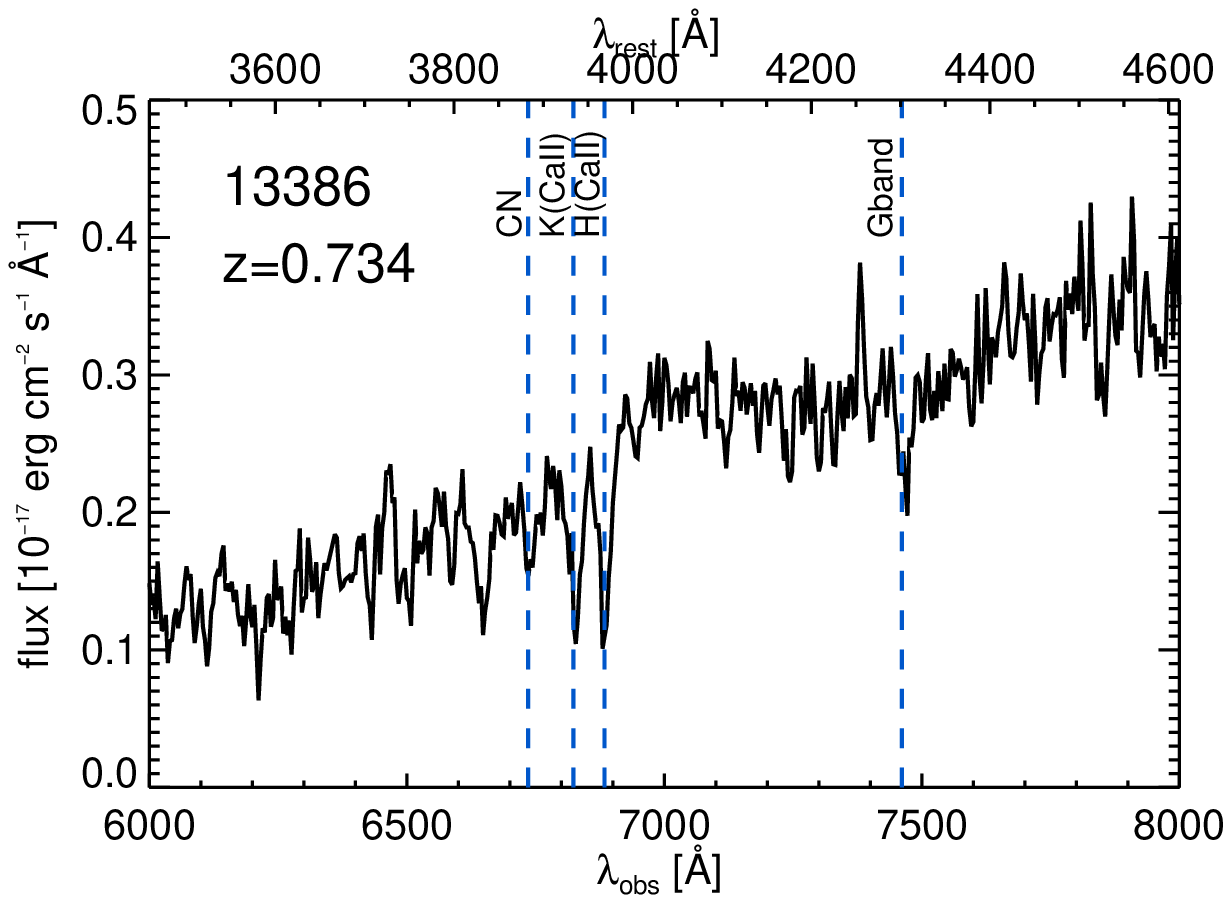}
\includegraphics[width=5.83cm]{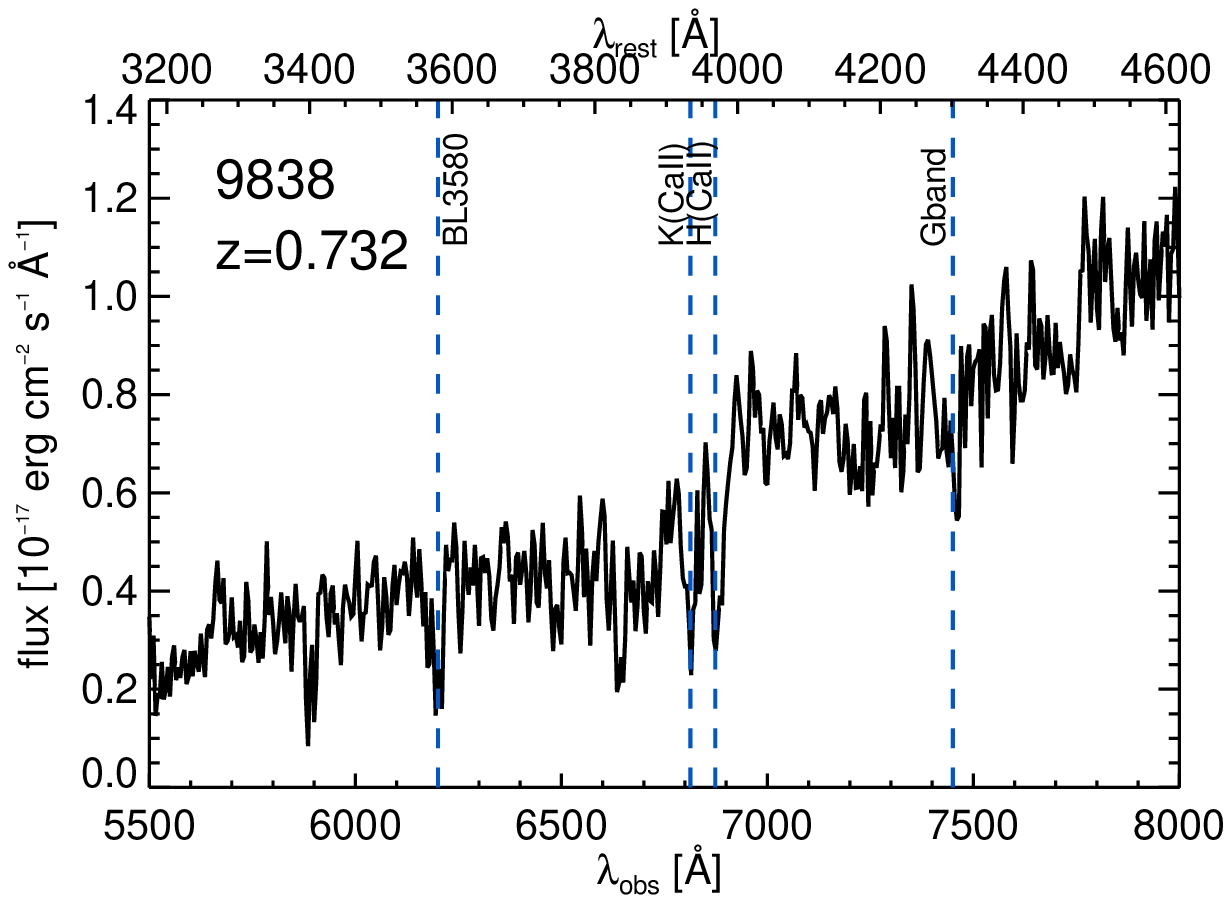}
\includegraphics[width=5.83cm]{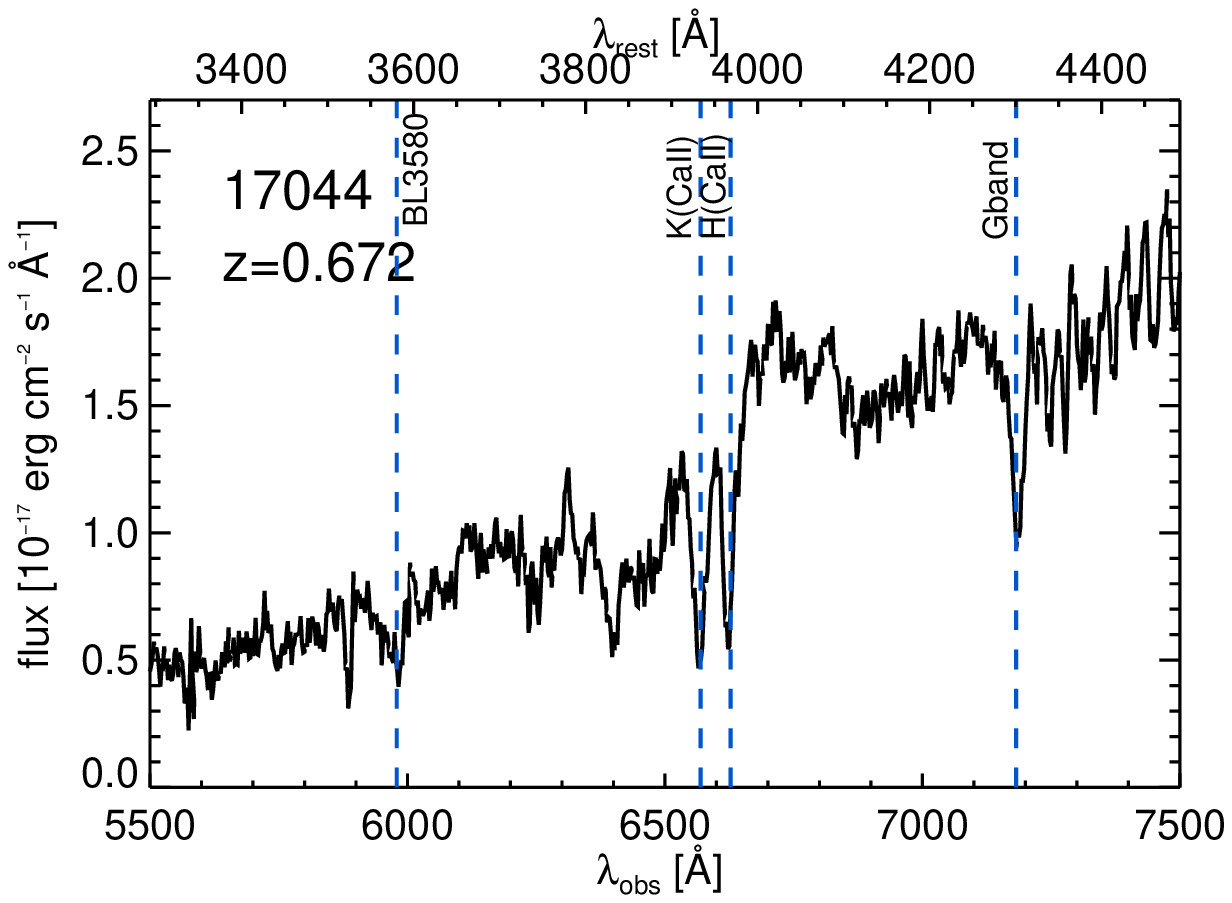}
\includegraphics[width=5.83cm]{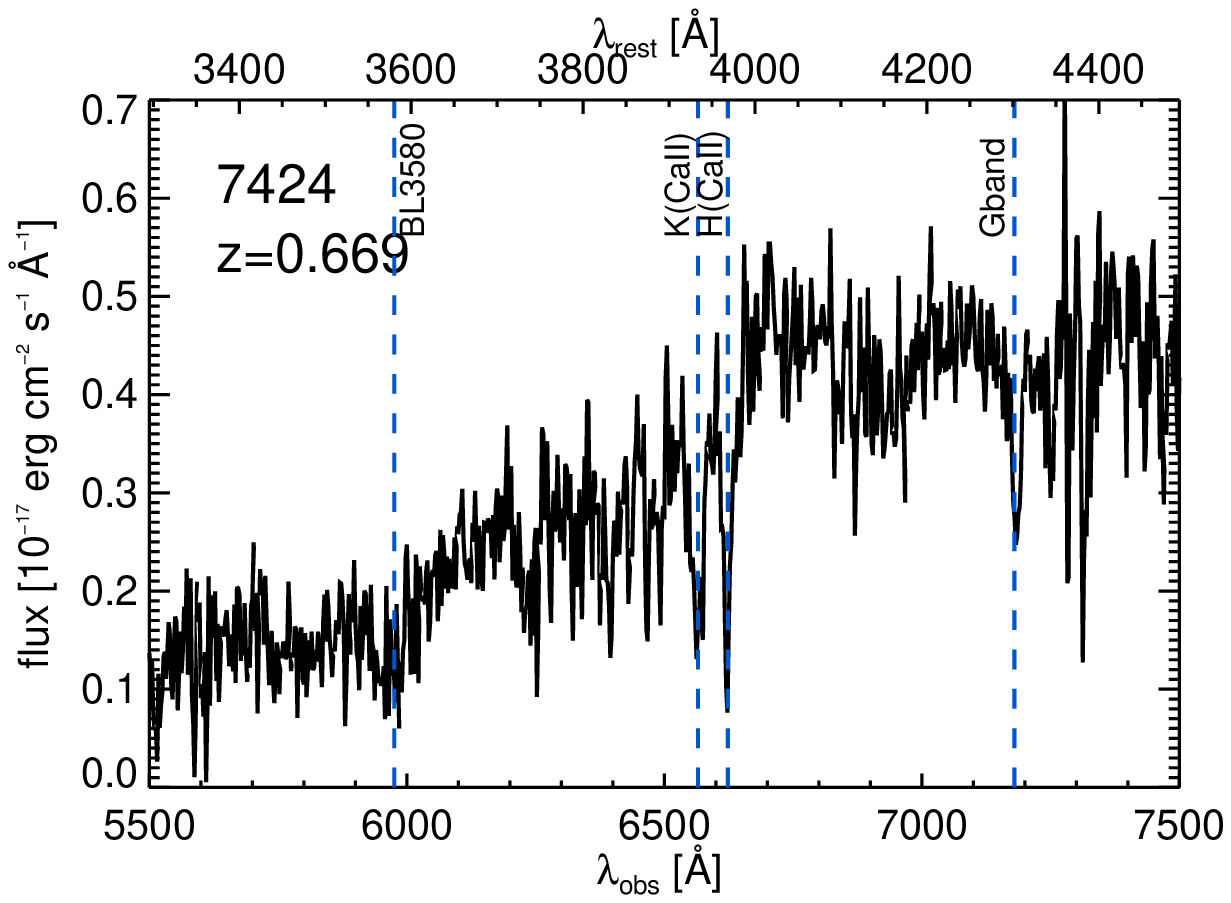}

\caption{Spectra of the sample galaxies with the main absorption lines in the region
of the Balmer/$4000$\AA $ $ break. Information about the spectroscopic data is shown in Table \ref{tab-spectradata}. Spectra are shown in bins spanning from $2.5$ to $8$
\AA.}
\label{spettrimono}
\end{figure*}

Reduced monodimensional spectra of the whole sample are shown in Figure \ref{spettrimono} together with the principal absorption lines in the region of 4000\AA $ $
rest frame.

\section{SPECTRAL INDEX MEASURES}
\label{indici}
As can be seen in Figure \ref{spettrimono}, many absorption lines typical of the stellar atmospheres can be identified in the optical spectra of our selected sample:
Balmer series, BL3580, CN, CaII(H,K), Gband.
We have taken advantage of the quality of our sample spectra of ETGs in order to analyse the main properties of their stellar populations, in particular the \emph{age} of
their stellar content.

The most important age-dependent spectrophotometric indices in the
observed optical region are: the $\Delta4000$ index, defined by \citet{bruzual83} and \citet{hamilton85} as the ratio between mean fluxes in two different wide spectral
ranges around the 4000\AA $ $ rest frame
\begin{equation}
\Delta4000=\frac{\bar{F_{\nu}}(\lambda_1 - \lambda_2)}{\bar{F_{\nu}}(\lambda_3 - \lambda_4)}
\end{equation}
with $\lambda_1-\lambda_2=[4050$\AA$-4250$\AA$]$, $\lambda_3-\lambda_4=[3750$\AA$-3950$\AA$]$;
and the H+K(Ca{\small\texttt{II}}) index, defined by \citet{rose85} as the ratio between minimum fluxes of the two CaII absorption lines H and K
\begin{equation}
\centering
H+K(Ca{\small\texttt{II}})=\frac{F_{min}(\lambda_3,\lambda_4)}{F_{min}(\lambda_1,\lambda_2)}
\end{equation}
with
$\lambda_1=3926$, $\lambda_2=3940$, $\lambda_3=3961$, $\lambda_4=3975$.
In particular, the CaII(H) absorption line is blended with the Balmer H$\epsilon$ at $\lambda=3969.65$\AA $ $ (see details in \citealt{longhetti99}).

The choice of measuring the values of these two spectrophotometric indices was supported by their intrinsic peculiarities.
In fact, these indices not only are strongly dependent on stellar population's age variations, mainly for ages less than $6$-$7$
Gyr which is about the age of Universe at $z\sim1$, but they also present precious differences which have made their combined measure very
interesting.

The $\Delta4000$ index provides a measure of the age of the global stellar population: in fact, according with its
definition, this index evaluates the ratio of the old and red stars flux with respect to the UV flux emission of the younger population.
Higher is its value, older is the stellar content. The effect of the metallicity goes in the same direction of ageing, but this
degeneracy is minimal for stellar populations younger than $6$-$7$ Gyr (the upper limit of the age of our high-$z$ galaxies sample).
On the other hand, the H+K(Ca{\small\texttt{II}}) index is deeply dependent on the presence of young stellar populations: in fact, the ratio between H
and K(CaII) lines would be constant if it was not for the blend of H(CaII) with Balmer H$\epsilon$, which actually determines the index
value. In particular, the effective ratio between H(CaII)+H$\epsilon$ and K(CaII) becomes $<1$ only for hot stars of type earlier than F5
which have strong Balmer series absorptions, with a minimum value for A-type Balmer-dominated stars \citep{rose85}. \\

We measured $\Delta4000$ and H+K(Ca{\small\texttt{II}}) indices on the spectra of our sample of ETGs, and results are reported in Table \ref{tab:indici}. It is worthy to note that the
measure of the H+K(Ca{\small\texttt{II}}) index is sensitive to the spectral resolution of the spectral data (see \citealt{longhetti99}), and we take this into account when comparing
measures with models.
Indeed, each measure performed on the spectra was corrected by the corresponding factor which brings
the index value as it was measured on a spectrum with the higher resolution of models. In order to obtain these correction factors, we have degraded the spectrum of a reference
template, with the original resolution full width
at half-maximum (FWHM)$=3$ \AA $ $ (in the spectral range $3000-5000$ \AA $ $ rest frame), to the lower resolutions of the observed spectra. We derived the correction factors by dividing the
index values measured on the degraded spectra by those measured on the reference template.
In particular, these correction factors decrease the index values up to about $5$ per cent of their measured values in the case of the lower resolution grism (R$=260$).

Furthermore, the broadening of the absorption lines due to the intrinsic velocity dispersion of galaxies affects the measure of this index. We thus corrected
the raw H+K(Ca{\small\texttt{II}}) index measures obtained on the five ETGs observed at the highest spectral resolution \citep{vdwel} for this effect, reporting them to the case of null
velocity dispersion.
We followed the same procedure as for the resolution correction: taking a reference template spectrum which has the same spectral resolution
of \citet{vdwel} data, we enlarged their spectral lines convolving them with a Gaussian curve with different $\sigma$ in order to simulate the effect of the velocity
dispersion. We measured the H+K(Ca{\small\texttt{II}}) index values for each value of the applied velocity dispersion and derived the correction factors to be applied to
the index measurements. The correction factors were found to be rather small and well within the error bars. In particular, for three
objects the correction was within $2$ per cent of the measured value, and for the other two objects it was about $4$ per cent.
For all the other $10$ ETGs of our sample, the lower spectral resolution of their observations prevents to appreciate the latter effect, being much
larger than the broadening caused by the resolution itself. Indeed, we have verified that for the medium-resolution spectra
\citep{popesso}, the corresponding correction factors, assuming a typical velocity dispersion of $250$ km/s, would be less than $1$ per cent of the measured index
values.

Errors have been computed from the uncertainties in the measures of fluxes and in the flux calibration.
In particular, we have estimated the values of S/N ratios in the two spectral regions involved in the index definitions (i.e. $3750-4000$\AA $ $ and $4050-4250$\AA)
which represent the statistical errors affecting the flux values.
On the other hand, we took into account also the flux calibration uncertainties derived from both the standard flux calibration and from the further correction
applied on the FORS-2 spectra which suffer for the shape distortion due to the CCD slit position (see Section \ref{datareduction}). Obviously, the H+K(Ca{\small\texttt{II}}) index, being the
flux ratio at two very close wavelengths, is almost unaffected by these flux calibration uncertainties. The final errors are shown in Table \ref{tab:indici}.

\begin{table}
 \centering
 \caption{Values of the measured H+K(Ca{\small\texttt{II}}) and $\Delta4000$ indices.}
 \begin{tabular}{lcc}
 \hline
 ID-MUSIC & H+K(Ca{\small\texttt{II}}) & $\Delta$4000\\
 \hline
 1192 & 0.75$\pm$0.07 & 1.80$\pm$0.04\\
 1382 & 1.14$\pm$0.07 & 2.15$\pm$0.07\\
 1950 & 0.97$\pm$0.23 & 2.05$\pm$0.1\\
 1837 & 1.49$\pm$0.10 & 2.16$\pm$0.3\\
 2694 & 1.07$\pm$0.16 & 1.87$\pm$0.06\\
 9066 & 1.34$\pm$0.27 & 1.65$\pm$0.10\\
 11539& 1.15$\pm$0.11 & 2.10$\pm$0.05\\
 10020& 1.57$\pm$0.23 & 1.84$\pm$0.06\\
 10960& 1.31$\pm$0.20 & 1.72$\pm$0.04\\
 11225& 0.91$\pm$0.13 & 1.81$\pm$0.03\\
 9792 & 1.15$\pm$0.13 & 2.09$\pm$0.04\\
 13386& 0.97$\pm$0.15 & 1.76$\pm$0.03\\
 9838 & 0.84$\pm$0.20 & 1.99$\pm$0.06\\
 17044& 1.16$\pm$0.12 & 2.05$\pm$0.02\\
 7424 & 0.58$\pm$0.23 & 1.81$\pm$0.07\\
 \hline
\end{tabular}
\label{tab:indici}
\end{table}

\section[]{Analysis}
\label{analisi}

The measured values of the indices have been
compared with the expectations of the spectro-photometric synthesis models (BC03) assuming a Chabrier IMF and an
exponentially declining star formation history with time-scale $\tau$ (with the same assumptions of the photometric analysis, see Section \ref{sample}, i.e.
$\tau=0.1-1$ Gyr and solar metallicity; afterwards we will discuss the effects of varying these assumptions).
Indeed, to start with the simpler evolutionary modelling of star formation typical of ETGs, we assumed that the bulk of the star formation
activity has happened in a
confined initial episode with small time-scale $\tau$ ($\tau<1$ Gyr, as it is the case of ETGs, see \citet{34etg}).
In this scenario, the stellar content of galaxies is
rather coeval, within $\tau$, so if this evolutionary modelling is appropriate for the real star formation history of ETGs, we expected that the stellar ages derived from the
values of the two indices would be consistent with each other.

\begin{figure}
\includegraphics[width=9cm]{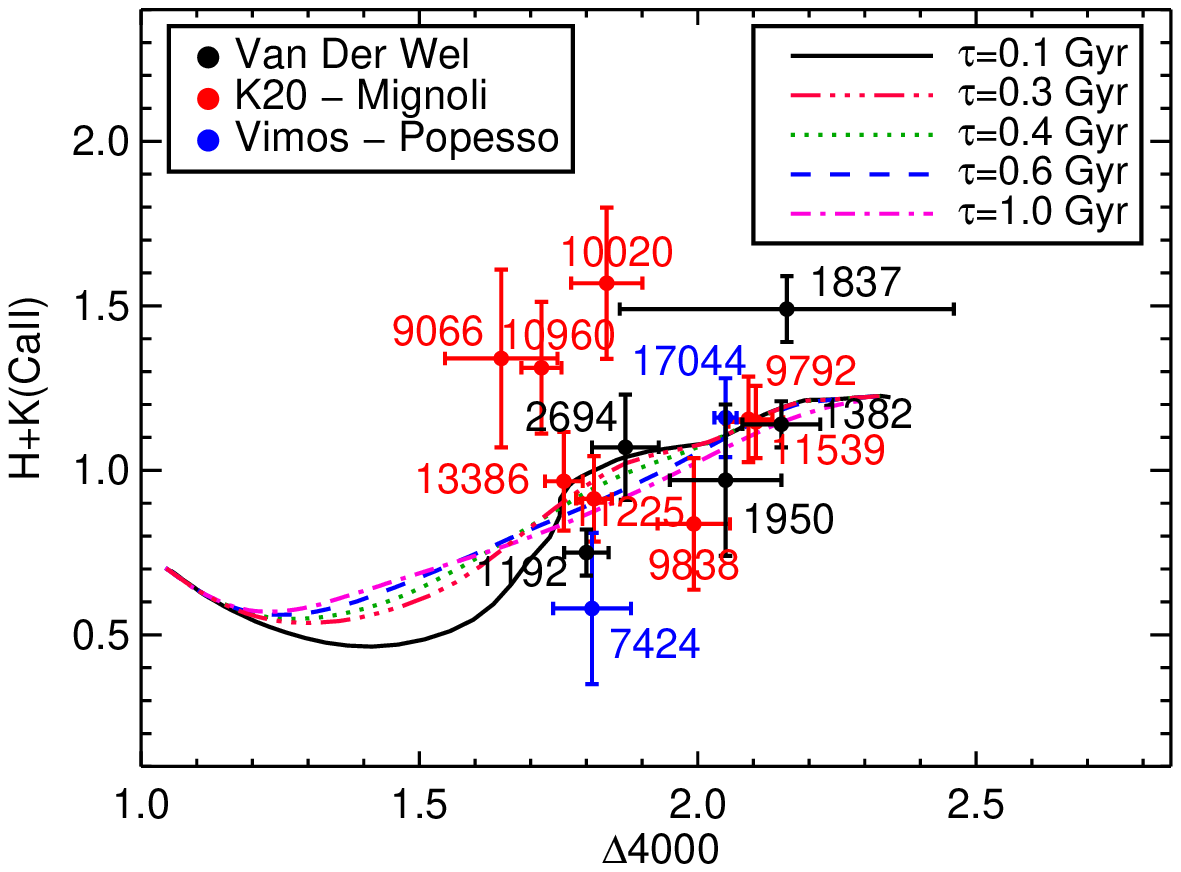}
\caption{Data versus model comparison in the index index plane: measured values of the indices $\Delta4000$ (\emph{x}-axis) and H+K(Ca{\small\texttt{II}}) (\emph{y}-axis) from the spectroscopic sample
	of \citet{vdwel} (black dots), \citet{mignoli} (red dots) and \citet{popesso} (blue dots), compared with the expectations of BC03 models at different ages
	(from left to right, from about $10^8$ to $10^{10}$ yr), fixed solar metallicity and for different \emph{star formation time-scale parameters} $\tau$
	(colored lines). With the exception of those from \citet{vdwel} sample, shown data points have H+K(Ca{\small\texttt{II}}) index values a bit underestimated (within error bars)
	with respect to those indicated by models due to the difference between the spectral resolutions of BC03 synthetic spectra (FWHM$=3$\AA) and that of the observed ones
	(see Table \ref{tab-spectradata}).}
\label{confrontomodelli}
\end{figure}

Figure \ref{confrontomodelli} shows the comparison between data and models in the $\Delta4000$ versus H+K(Ca{\small\texttt{II}}) plane. 
Black dots are galaxies from the spectroscopic sample of \citet{vdwel}, red ones from \citet{mignoli} and blue ones from \citet{popesso}. Lines
represent the trends of the values of the two indices of synthetic models at different ages, from younger ages ($10^8$ yr) at the bottom left to older ages ($10^{10}$ yr)
at the upper right (at fixed solar metallicity), and their colours show different star formation histories determined by the time-scale parameter $\tau$.
From Figure \ref{confrontomodelli} it is possible to notice that: \emph{i)} data points cover an extended area in the index index plane which corresponds to stellar
populations older than $\sim2.5$ Gyr; \emph{ii)} different $\tau$-models (coloured lines) are confined in a rather thin strip,
while measured data are more widespread. While the majority of the data points are in
agreement with models, we get the evidence that for $6$ out of $15$ objects both indices deviate at more than $1\sigma$ from models. In particular, starting from
the measured values of the $\Delta4000$ index, $7$ out of $15$ data
points have: either too low values of the H+K(Ca{\small\texttt{II}}) index (IDs: $1192$, $7424$ and $9838$), suggesting possible stellar populations younger
than what pointed out from the $\Delta4000$ index (see Section \ref{indici}), or so high values of the H+K(Ca{\small\texttt{II}}) index (H+K(Ca{\small\texttt{II}})$>1.2$, IDs:
$9066$, $10960$, $10020$ and $1837$) that they cannot find any correspondence with models within $2\sigma$.
It is important to notice that our sample of galaxies is not complete, so these
percentages are just evidence that at least some of these high-redshift ETGs have a less homogeneous stellar population with respect to what is expected in a rigorous passive evolution.

\begin{figure}
\includegraphics[width=9cm]{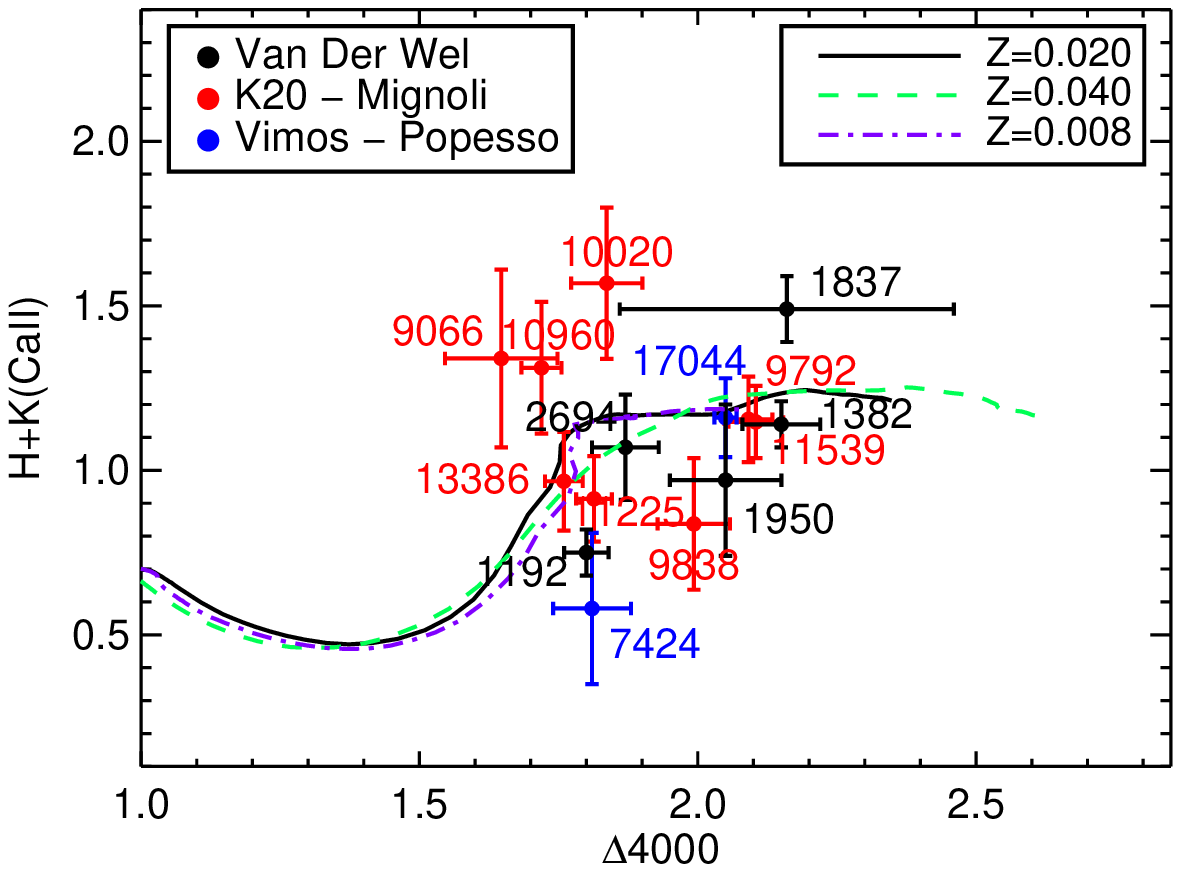}
\caption{Data versus models comparison in the index index plane: measured values of the indices $\Delta4000$ (\emph{x}-axis) and H+K(Ca{\small\texttt{II}}) (\emph{y}-axis) from the spectroscopic sample of
	\citet{vdwel} (black dots), \citet{mignoli} (red dots) and \citet{popesso} (blue dots), compared with the expectations of BC03 models at different ages
	(from left to right, from about $10^8$ to $10^{10}$ yr), fixed star formation time-scale parameter $\tau$ and for different \emph{metallicities}: Z $=0.008$,
	Z $=0.020$ and Z $=0.040$ (coloured lines). As in Figure \ref{confrontomodelli}, with the exception of those from \citet{vdwel} sample, shown data points have
	H+K(Ca{\small\texttt{II}}) index values a bit underestimated (within error bars) with respect to those indicated by models due to the difference between the spectral resolutions
	of BC03 synthetic spectra (FWHM$=3$\AA) and that of the observed ones (see Table \ref{tab-spectradata}).}
\label{metallicity}
\end{figure}

Points that are not consistent with the proposed models at different star formation time-scales cannot be explained by using synthetic models with different
\emph{metallicities}. As an example, in Figure \ref{metallicity} the same diagram of Figure \ref{confrontomodelli} is shown, but in this case the different coloured lines
represent models with three distinct metallicities (Z $=0.008$, Z $=0.020$ and Z $=0.040$), at fixed time-scale parameter ($\tau=0.1$ Gyr). From Figure \ref{metallicity} it
is clear that the not consistent points still remain without a satisfactory explanation on the basis of single-component star formation histories even at different metallicities.

Adding some amounts of dust (A$_v>0.5$ mag) in the single component models would marginally help in explaining those data points which have low values of the H+K(Ca{\small\texttt{II}})
index (IDs: $1192$, $7424$ and $9838$). However, even this addition is not able to reproduce the index values of four objects with high H+K(Ca{\small\texttt{II}}) (IDs: $9066$, $10960$, $10020$ and
$1837$) because its main effect is to little increase the values of the $\Delta4000$ index.
Moreover, this choice should be supported by a physical motivation able to justify these quantities of dust in systems where globally the star formation
activity should be quenched (age $>>\tau$).

We have also checked that the values of the two measured indices on synthetic models do not depend on the spectrophotometric models adopted in the analysis: comparing index
values obtained from BC03 models with those obtained with the latest version of the same code (Charlot $\&$ Bruzual, in preparation), we do not find any difference,
while the comparison with \citet{ma11} models reveals small differences but well within the observational errors.\\

\vskip 0.3cm
\noindent

The comparison between the measured indices and models has thus suggested that the star formation histories of these galaxies must have been more complex, eventually
involving the presence of later star formation events superimposed on a rather old bulk of stars, causing a less homogeneous population than predicted by simple short
$\tau$ models.

A further confirmation of this evidence comes from the discrepancy we found among ages derived from the values of the two indices and those extracted from the SED fitting
analysis, as shown in Table \ref{tab:ages}: the $\Delta4000$ index points towards generally more evolved stellar populations. Ages deduced from the index values and
reported in Table \ref{tab:ages} have been derived assuming BC03 models, solar metallicity and $\tau=0.1$ Gyr. Ages deriving from indices and from SED fitting are
two different but complementary tools to study the stellar content of galaxies: the spectroscopic analysis is based on a small portion of the SED but it is able to
give important details on their recent star formation history, while the photometric analysis can provide mean stellar parameters dealing with the whole SED. This first result thus suggests that in general the SED analysis is missing small pieces of information, revealed instead by the indices.

\begin{table*}
 \centering
 \begin{minipage}{140mm}
 \caption{Age estimates derived from indices and photometric analysis (Gyr) obtained from the comparison with synthetic stellar populations models with $\tau=0.1$ Gyr and solar metallicity.}
 \begin{tabular}{cccc}
 \hline
 ID-MUSIC & Age from H+K(Ca{\small\texttt{II}}) & Age from $\Delta$4000 & Age from photometry\\
          & (Gyr)              & (Gyr)                 & (Gyr)     \\
 \hline
 1192 & 1$\pm$0.1              & 1.9$\pm$0.5           & 0.7\\
 1382 & 4.5$\pm$1.5            & 5.5$\pm$1.5           & 1.4\\
 1950 & 1.7$\pm$2              & 4$\pm$1               & 0.7\\
 1837 & -                      & 5.75$\pm$3            & 0.9\\
 2694 & 3$\pm$1.2              & 2.4$\pm$0.4           & 1.3\\
 9066 & -                      & 0.9$\pm$0.4           & 1.3\\
 11539 & 4.7$\pm$1.5           & 4.75$\pm$0.5          & 1.3\\
 10020 & -                     & 2.2$\pm$0.5           & 1.3\\
 10960 & -                     & 1.1$\pm$0.3           & 1.4\\
 11225 & 1.2$\pm$0.5           & 1.9$\pm$0.4           & 1.4\\
 9792  & 4.7$\pm$1.5           & 4.7$\pm$0.9           & 2.7\\
 13386 & 1.4$\pm$0.6           & 1.5$\pm$0.4           & 0.6\\
 9838  & 0.9$\pm$0.3           & 3.1$\pm$0.4           & 2.3\\
 17044 & 1.9$\pm$1             & 3.75$\pm$0.4          & 2.3\\
 7424  & 0.7$\pm$0.4           & 1.9$\pm$0.5           & 1.4\\
 \hline
 \end{tabular}
 \end{minipage}
 \label{tab:ages}
\end{table*}

Moreover, for many objects ($8$ out of $15$) the ages derived from the $\Delta4000$ index tend to differ by more than $1\sigma$ from those derived from the H+K(Ca{\small\texttt{II}})
index. In particular, in one case they differ by more than $2\sigma$ (ID-$9838$). Furthermore, even if possible, it appears unlikely that the age
inconsistency arisen from the two indices is due to measurement errors. Indeed, it would mean introducing a kind of systematicity which should always decrease the values of
age derived from the H+K(Ca{\small\texttt{II}}) index and at the same time always increase those derived from the $\Delta4000$ index.
In fact, ages from the $\Delta4000$
index are older than those indicated from H+K(Ca{\small\texttt{II}}) index (only for one object there is the opposite trend), in agreement with \citet{onodera} who found similar
results for passive galaxies at $z>1.4$. Furthermore, in some cases, ID-2694,
ID-9066, ID-10020 and ID-10960 as displayed in Table \ref{tab:ages}, H+K(Ca{\small\texttt{II}}) values do not find any age correspondence with models, as visible in Figure
\ref{confrontomodelli}. As discussed in Section \ref{indici}, due to the nature of the two selected indices, we expect dissimilarity between them if the galaxy stellar
content reveals non-homogeneous properties, especially age. For this reason, having found such a discrepancy in most cases, we started a new analysis of the star
formation histories of our sample galaxies assuming that a quite recent star-forming event has been superimposed on an otherwise homogeneous and old stellar population.

As a pure exercise, we have performed the analysis based on the assumption of a double age stellar component for all the galaxies of the sample. The real convenience of
introducing the second component, as discussed in the following section, is found only for those objects for which the age values derived from the two indices
were not in agreement. And this is the case for $9$ out of $15$ galaxies. For the remaining galaxies, values of $\tau \geq 0.1$ Gyr are
sufficient to explain the inhomogeneity of their stellar content even though with some corrections to the mean ages derived from the SED fitting analysis.


\subsection{Double-component analysis}
\label{sec-double}

To better reproduce the observed stellar populations properties of our sample galaxies and in particular to reproduce the measured indices, we thus hypothesized the
presence of \emph{composite} stellar populations, assuming that after the initial bulk of star formation, additional events have occurred also in the following few Gyr.
For the sake of simplicity, we started introducing only \emph{double} stellar components; we cannot exclude that a more refined analysis which involves more than two
spectral index measurements on higher quality spectra could eventually find the presence of triple and over stellar populations.

\begin{figure*}
\includegraphics[width=8.8cm]{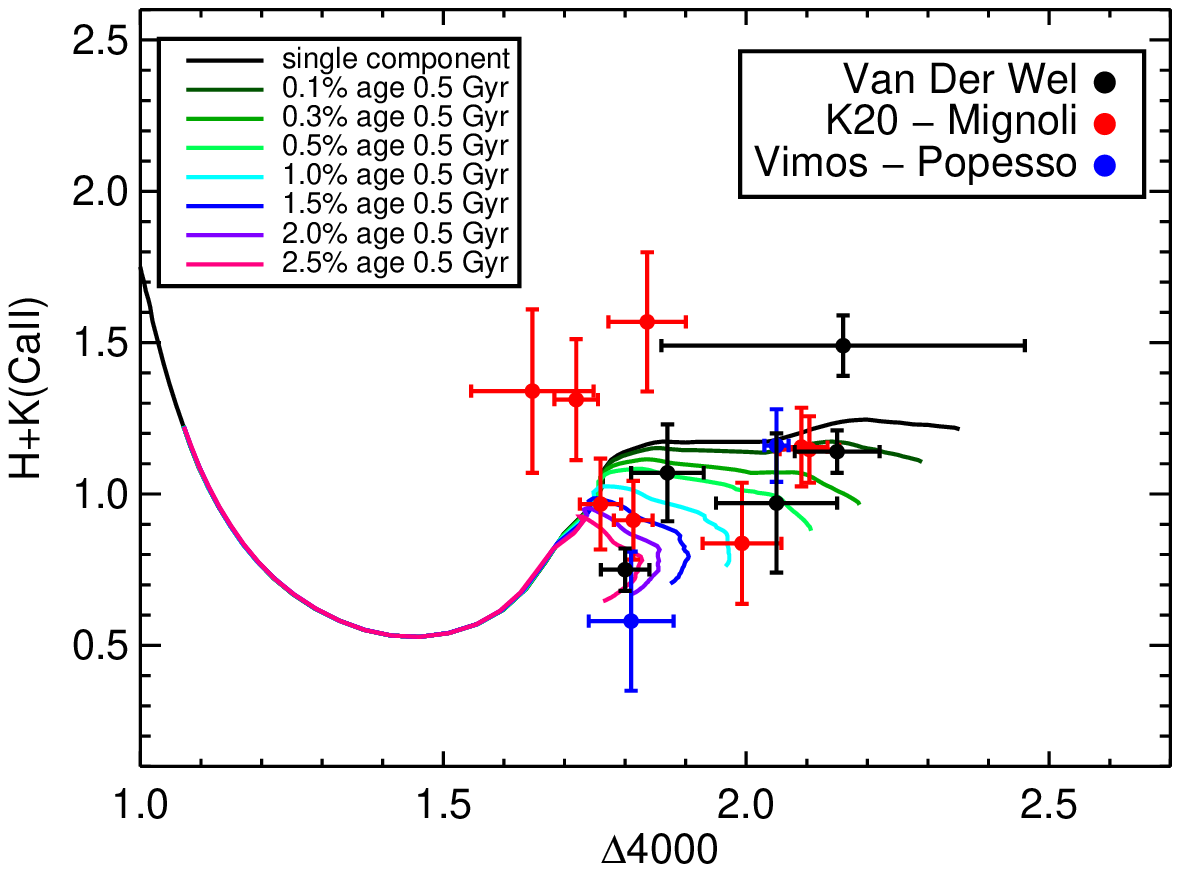}
\includegraphics[width=8.8cm]{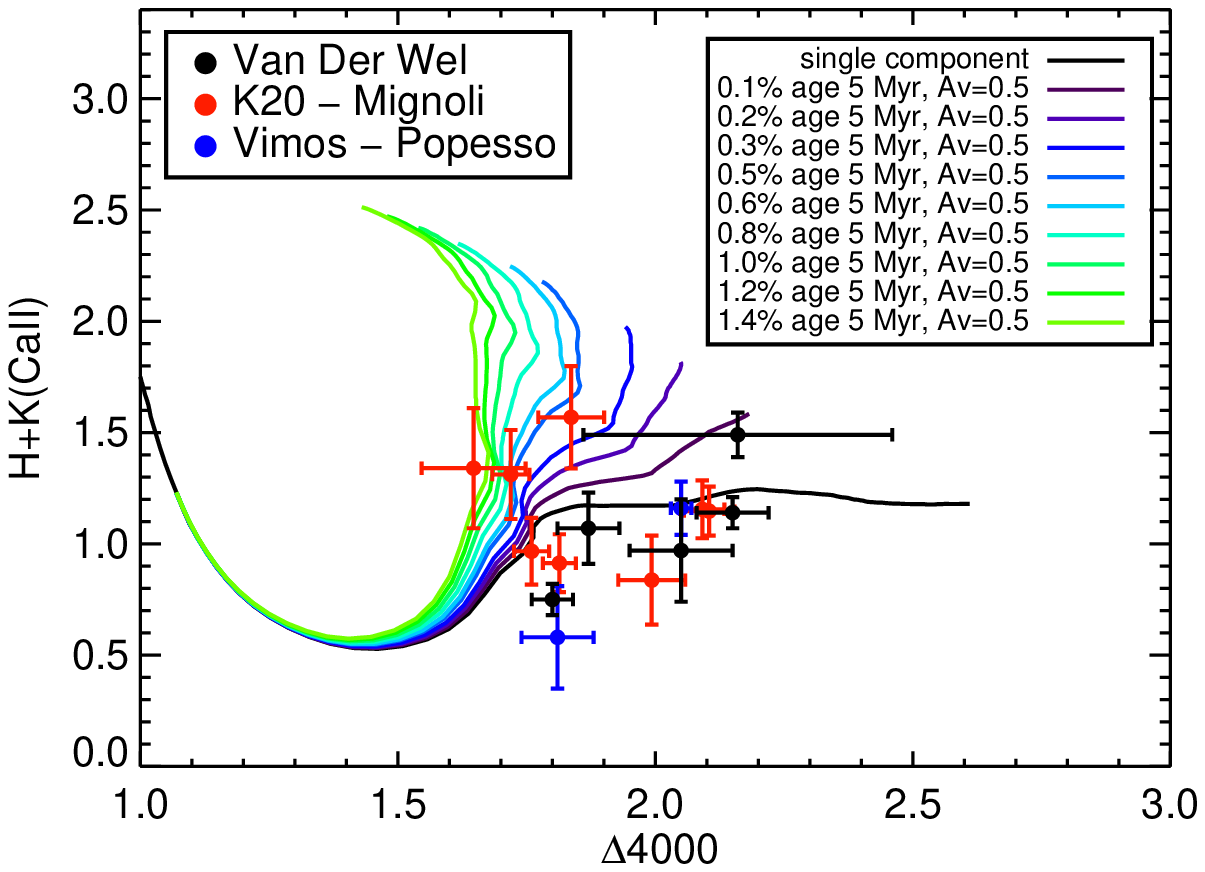}
\caption{Index index plane: comparison with \emph{double-component} models. Left-hand panel: observed points are the same as in Figure \ref{confrontomodelli}; lines represent the
values of the indices on BC03 synthetic models composed of a main component whose age increases from left to right (from about $10^8$ to $10^{10}$ yr), and by small
mass fraction of a younger stellar component with age $0.5$ Gyr. The fraction of the secondary component increases from $0.1\%$ (dark green line) to $2.5\%$ (magenta
line). Right-hand panel: same as the left-hand panel; here the small and younger component has age $5$ Myr and it has been reddened with a typical dust extinction of 0.5 mag.
Mass fractions span from $0.1\%$ (violet line) to $1.4\%$ (green line). BC03 models are with fixed solar metallicity and $\tau=0.1$ Gyr. As in Figure
\ref{confrontomodelli} and \ref{metallicity}, with the exception of those from \citet{vdwel} sample, shown data points have H+K(Ca{\small\texttt{II}}) index values a bit underestimated
(within error bars) with respect to those indicated by models due to the difference between the spectral resolutions of BC03 synthetic spectra (FWHM$=3$\AA) and that of
the observed ones (see Table \ref{tab-spectradata}). Both plots show how those data points which were not consistent with \emph{single-component} models can be explained.}
\label{confrontodouble}
\end{figure*}

In Figure \ref{confrontodouble} (left-hand panel), the comparison of the measured indices with those obtained on synthetic double-component models is shown: models represent
galaxies for which the bulk of the stellar mass belongs to a population whose age increases from left to right (from about $10^8$ to $10^{10}$ yr) in the index index
plane, and small mass fractions are taken into account by a younger component with fixed age $0.5$ Gyr. Coloured lines represent the increasing intensity of the secondary
burst, i.e. the fraction of the younger component, from the right (dark green line) with a fraction of $0.1$ per cent to the left (magenta line) with $2.5$ per cent.
From this plot, it is clear that with only small mass fractions of a superimposed younger population, observed points with low H+K(Ca{\small\texttt{II}}) index values can be fully
explained.

\begin{figure}
\includegraphics[width=9cm]{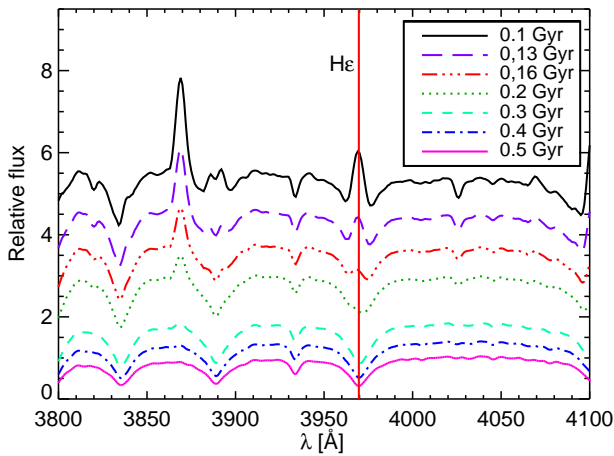}
\caption{Synthetic spectra with Balmer H$\epsilon$ ($\lambda=3969.65$ \AA $ $, red vertical line) emission line in models at different stellar population ages.
         Ages decrease from $0.5$ Gyr (magenta line) to $0.1$ Gyr (black line).}
\label{emissioni}
\end{figure}
	
\begin{figure}
\includegraphics[width=9cm]{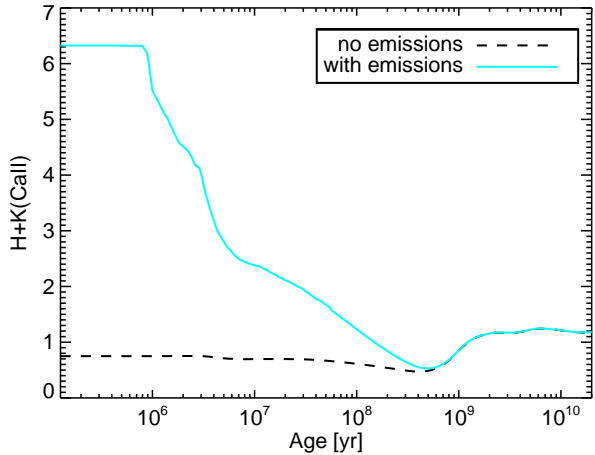}
\caption{Trend of the H+K(Ca{\small\texttt{II}}) index with age measured on BC03 models with Balmer lines emission (cyan solid line) and without emission (black dashed
line). Differences between two models become important only for stellar population with age $<1$ Gyr.}
\label{diffemissioni}
\end{figure}

A similar explanation can be possibly advanced also for points with higher H+K(Ca{\small\texttt{II}}) index values, but with the addition of a further
ingredient.
Considering the definition of the H+K(Ca{\small\texttt{II}}) index, as discussed in Section \ref{indici}, high index values would mean that the ``H'' absorption line must be very
weak with respect to the K(CaII) line. We then hypothesized that the Balmer H$\epsilon$ line blended with the H(CaII) line is in \emph{emission} rather than in
absorption. As shown in Figure \ref{emissioni} in BC03 models, for stellar populations younger than $<0.2$ Gyr, Balmer H$\epsilon$ emission is dominant over the absorption.
So that the presence of a small mass fraction of very young stellar population (i.e., very recent small burst) can be able to fill the depth of the whole H line,
thus increasing the value of the H+K(Ca{\small\texttt{II}}) index. We then introduced in our analysis BC03 models which include the nebular emission lines (see \citealt{charlot2001}).
In fact, as shown in Figure \ref{diffemissioni}, differences between models with emission (cyan solid line) and
without emission (black dashed line) are evident for ages smaller than $1$ Gyr; for older ages these models coincide.
As an example of our analysis, in Figure \ref{confrontodouble} (right-hand panel) the comparison of model prediction with the observed data in the case of double age component
models is shown for a starburst age of $5$ Myr. Given the very young age of the secondary component, we took into account the possible dust extinction that should be
acting around it, assuming A$_v=0.5$; we specify that the introduction of dust affects only the $\Delta4000$ index, increasing its value in the same way as age. The intensity of the star formation
burst increases from right (violet line), with a mass fraction of $0.1$ per cent, to left (green line), with a mass fraction of $1.4$ per cent. Both panels of Figure \ref{confrontodouble} are
useful examples that show how the suggested presence of small mass fraction of younger stellar component is able to explain the observed data. Age and percentage of mass
involved in the secondary component (i.e., age and strength of the recent burst) can be varied to exactly match each single observation. Below we present the detailed
analysis of the whole sample of galaxies and the obtained results.

\vskip 0.3cm
\noindent

Each galaxy has been analysed by means of a code that, spanning all the double-component possible models, extracts best-fitting solutions, following the analogue of the
one-component photometric analysis. Assuming the BC03 models, including possible emission nebular components, with fixed $\tau=0.1$ Gyr and solar metallicity,
the code builds all the possible double-component combinations of spectra varying ages, mass fractions and dust extinction of the two stellar components. A maximum
age limit has been imposed equal to the age of the Universe at the redshift of each galaxy. Recent burst strengths (i.e. the mass percentage involved in the young component)
have been considered by means of $0.05$ per cent steps. Dust extinction has been applied both on the whole double-component spectrum and only on the younger component,
in parallel distinct analysis, with a $0.1$ mag pass.

From the generated composed synthetic spectra our code selects the solutions which are consistent with the measured values of the two spectrophotometric indices within their
observational errors. After this selection, the code starts the $\chi^2$ minimization process of the composite synthetic SED on the observed photometric data. In
this way not only the optical band is considered, but the whole spectral distribution.

The results of the analysis are shown in Table \ref{tab:results}. Mass fraction values expressed in Table \ref{tab:results} are indicative of the amount of
mass for the solutions with lower $\chi^2$ values. The sample objects are well described by two-component models, revealing the consistence of the found solutions with
the presence of a younger stellar population for almost all the galaxies of the sample.
Moreover, thanks to the procedure applied in our code, all the composite spectra
are fully consistent with the measured indices, contrary to the single-component analysis from which this happens only for a couple of objects. Dust extinction in
general becomes important, i.e. A$_v>2$, only for objects whose young component is younger than the star formation time-scale parameter $\tau=0.1$ Gyr, that means that
the star formation is still ongoing.

The improvement in terms of probability P($\chi^2$) of the obtained double-component solutions with respect to the single component ones (with $\tau=0.1$ Gyr, see
Table \ref{tab:ages}) is illustrated in Table \ref{tab:chi2}, where we report the comparison of the reduced-$\chi^2$ ($\chi^2_{\nu}$ ). The
estimation of the $\chi^2_{\nu}$ includes both the contribution of the $14$ photometric points which constitute the global SED and the values of the two analysed spectral
indices. It is worth to remark that in this part of the analysis, for consistency, we have compared the double-component models with the single component ones obtained with fixed
small $\tau=0.1$ Gyr. Accordingly, the $\chi^2_{\nu}$ values and the corresponding probabilities shown in Table \ref{tab:chi2} are referred to these particular SED fitting
solutions.

As can be seen from Table \ref{tab:chi2}, the double-component solutions are preferable over the single component ones for almost all the
galaxies of the sample. More in details, for $9$ out of $15$ galaxies ($60$ per cent of the sample) (IDs: 1192, 1950, 1837, 9066, 10020, 10960, 9838, 17044 and 7424), which have the age values derived from indices
not in agreement with each other (see Table \ref{tab:ages}), the double-component solutions with age differences $> 1.5$ Gyr are clearly more
representative of their inhomogeneous stellar contents. For the other six objects, even if providing statistically more probable solutions for four out of six cases,
the double-component analysis appears not significantly necessary. Indeed,  for these six galaxies the ages derived from the two indices result all in agreement with each other pointing
to the same generally older age with respect to the ages extracted from the SED fitting analysis (see Table \ref{tab:ages}). In particular, for three galaxies (IDs: 1382, 11539 and
9792), the double-component solutions are statistically more probable actually
just because the bulk of the mass ($>98$ per cent) matches the age indicated by the two indices (compare Table \ref{tab:ages} with Table \ref{tab:results}), while the single-component
solutions miss this important information. Only for the remaining $3$ cases (IDs: 2694, 11225 and 13386) both indices and $\tau
\geq0.1$ Gyr SED fitting analysis point to the same age, directly revealing a rather homogeneous stellar population.

Summarizing, the double-component analysis shows that in $60$ per cent of the galaxies of our sample there are two stellar components whose presence explains the different
ages derived from the index values. Furthermore, thanks again to the index values, this analysis has confirmed or corrected the ages of the SED fitting solutions for the remaining
objects.


\begin{table*}
 \centering
 \begin{minipage}{140mm}
 \caption{Comparison of the reduced-$\chi^2$ and corresponding probabilities between the single and double-component
 solutions. The $\chi^2_{\nu}$ estimations include the comparison among the photometric points and the analysed spectral indices. }
 \label{tab:chi2}
 \begin{tabular}{lcccc}
 \hline
 ID   &  $\chi^2_{\nu}$ single component  & Associated probability  & $\chi^2_{\nu}$ double component  & Associated probability \\
      &                                   & ($\%$)                  &                                  &  ($\%$) \\
 \hline
 1192  & 5.01 & 0.7 &1.14 & 32 \\
 1382  & 13.6 & 0   &3.7  & 16 \\
 1950  & 7.5  & 0.06&2.94 & 5  \\
 1837  & 19.94& 0   &1.31 & 27 \\
 2694  & 1.56 & 21  &1.8  & 16 \\
 9066  & 3.31 & 4   &0.66 & 52 \\
 11539 & 12.66& 0   &1.16 & 31 \\
 10020 & 5.02 & 0.7 &0.844& 43 \\
 10960 & 3.85 & 2   &1.74 & 17 \\
 11225 & 0.59 & 55  &0.77 & 46 \\
 9792  & 9.05 & 0.01&1.64 & 19 \\
 13386 & 4.89 & 0.7 &1.16 & 31 \\
 9838  & 3.43 & 3   &1.26 & 28 \\
 17044 & 9.87 & 0.01&1.31 & 27 \\
 7424  & 2.038& 13  &1.01 & 36 \\
 \hline
 \end{tabular}
 \end{minipage}

\end{table*}


\begin{table*}
 \centering
 \begin{minipage}{160mm}
 \caption{Results obtained from the double-component analysis on the sample of galaxies extracted from the composition of BC03 models, with fixed $\tau=0.1$ Gyr and
 solar metallicity. Mass fractions are the indicative values for the solutions with lower $\chi^2$ values.}
 \label{tab:results}
 \begin{tabular}{lccccccc}
 \hline
 ID & Age young component & Mass fraction & Age old component & Mass fraction & A$_v$ & Age differences & $z_{form}$ bulk\\
    &(Gyr)                &   ($\%$)      &(Gyr)              &   ($\%$)      & (mag) &(Gyr)            &\\
 \hline
 1192 & 0.64 & 6.5  & 2.4 & 93.5  & 0   & 1.76     &2.15\\
 1382 & 0.1  & 0.05 & 5   & 99.95 & 0.9$^a$ & 4.9     &6.1\\
 1950 & 0.45 & 1    & 5.5 & 99    & 0   & 5.05     &$>$6\\
 1837 & 0.015& 0.95 & 5.5 & 99.05 & 2.8$^a$ & 5.485   &$>$6\\
 2694 & 0.5  & 0.65 & 2.3 & 99.35 & 0.1 & 1.8      &2.2\\
 9066 & 0.006& 0.75 & 1.7 & 99.25 & 2.8$^a$ & 1.694   &1.95\\
 11539 & 0.3 & 0.1  & 4.75& 99.9  & 0.2$^a$ & 4.45    &$>$6\\
 10020 & 0.03 & 1.5 & 2.6 & 98.5  & 2.3$^a$ & 2.57    &1.45\\
 10960 & 0.015 & 1.65 & 2.3 & 98.35 & 2.6$^a$ & 2.285 &1.32\\
 11225 & 0.9 & 20   & 6.25 & 80   & 0.2 & 5.35     &$>$6\\
 9792  & 0.8 & 1.65 & 4.75 & 98.35& 0.3 & 3.95     &2.86\\
 13386 & 0.1 & 0.1  & 1    & 99.9 & 0.3 & 0.9      &1\\
 9838  & 0.5 & 2.6  & 4    & 97.4 & 0.8$^a$ & 3.5     &2.16\\
 17044 & 1   & 5.3  & 7    & 96.5 & 0.2 & 6        &$>$6\\
 7424  & 0.7 & 16.7 & 6.25 & 83.3 & 0.5$^a$ & 5.55    &5.5\\
 \hline
 \end{tabular}
 $^a$A$_v$ applied only on the young component.
 \end{minipage}

\end{table*}

\begin{figure*}
\includegraphics[width=8.8cm]{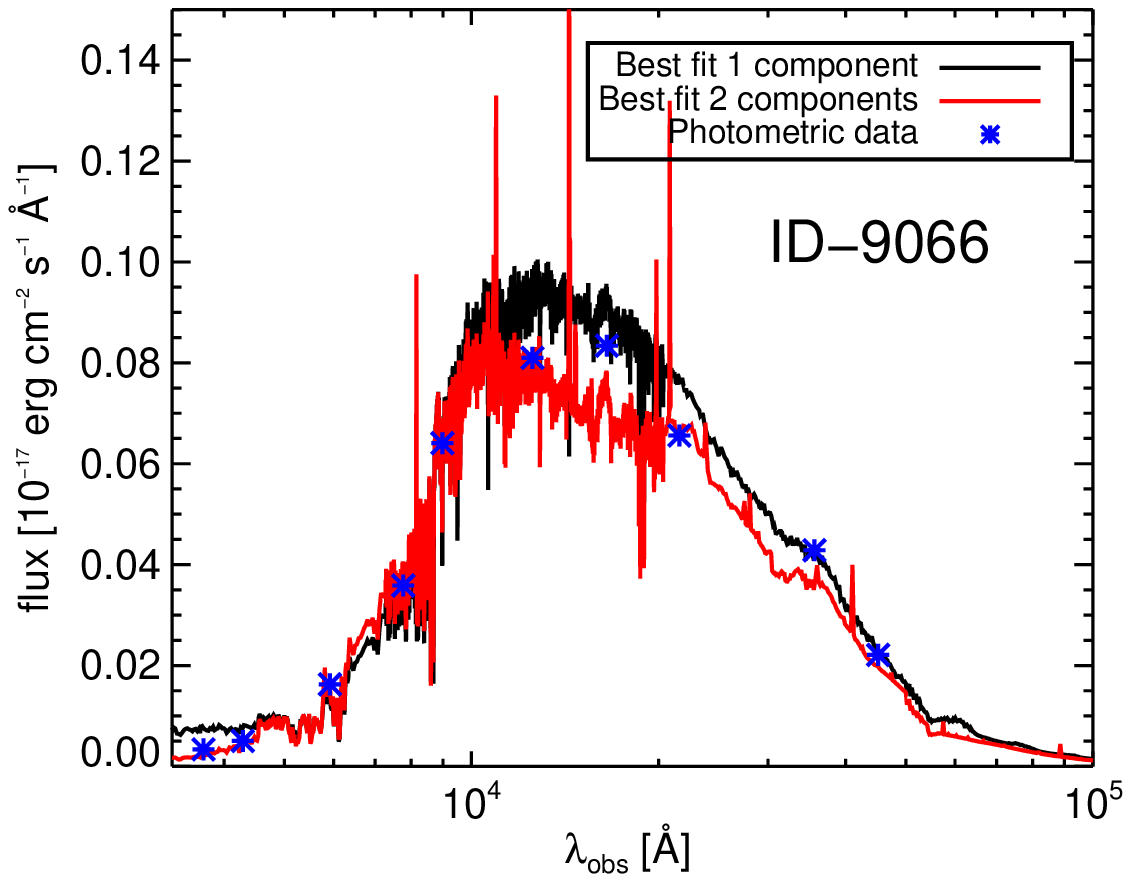}
\includegraphics[width=8.8cm]{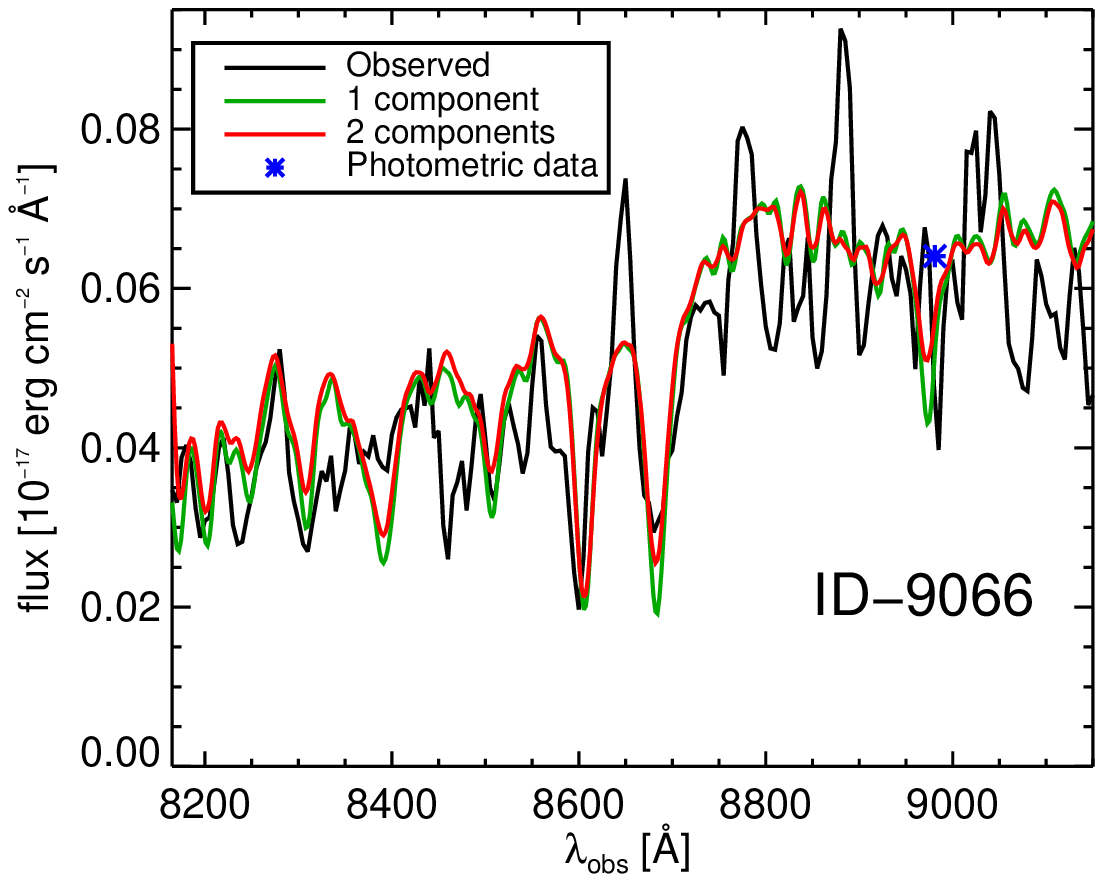}
\includegraphics[width=8.8cm]{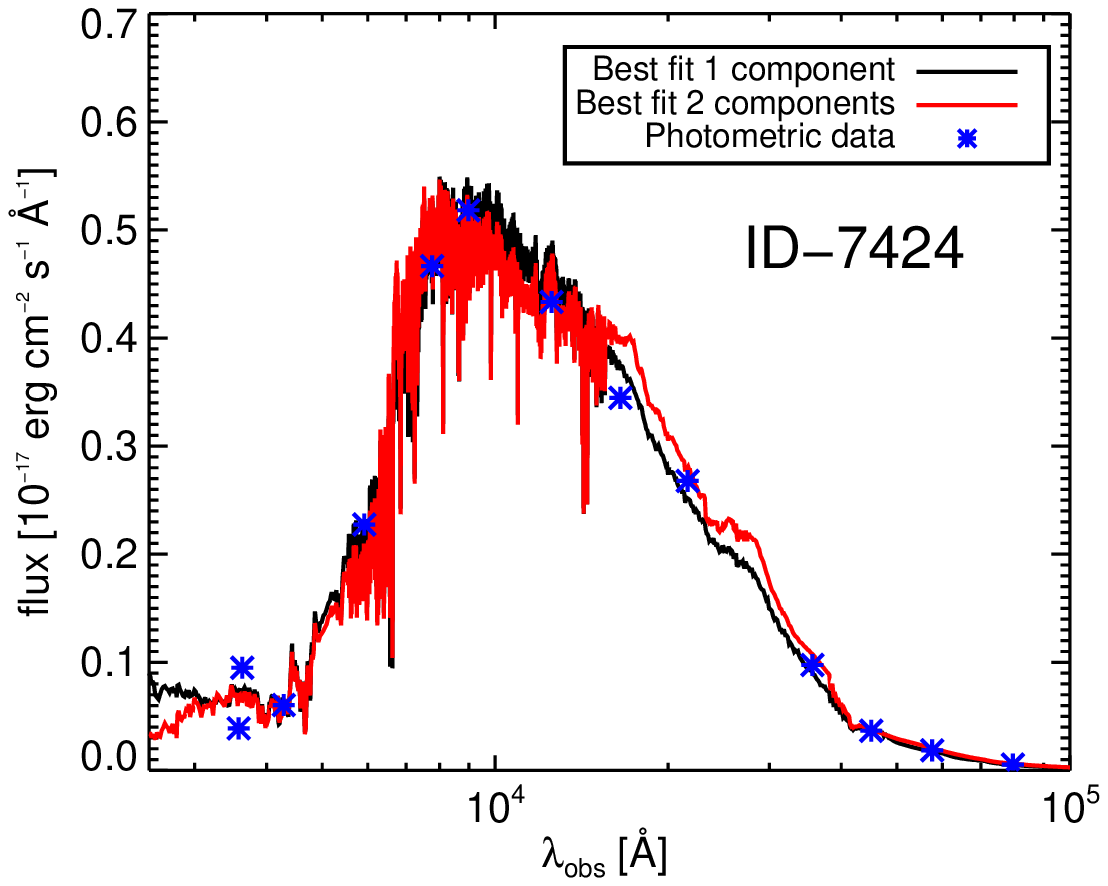}
\includegraphics[width=8.8cm]{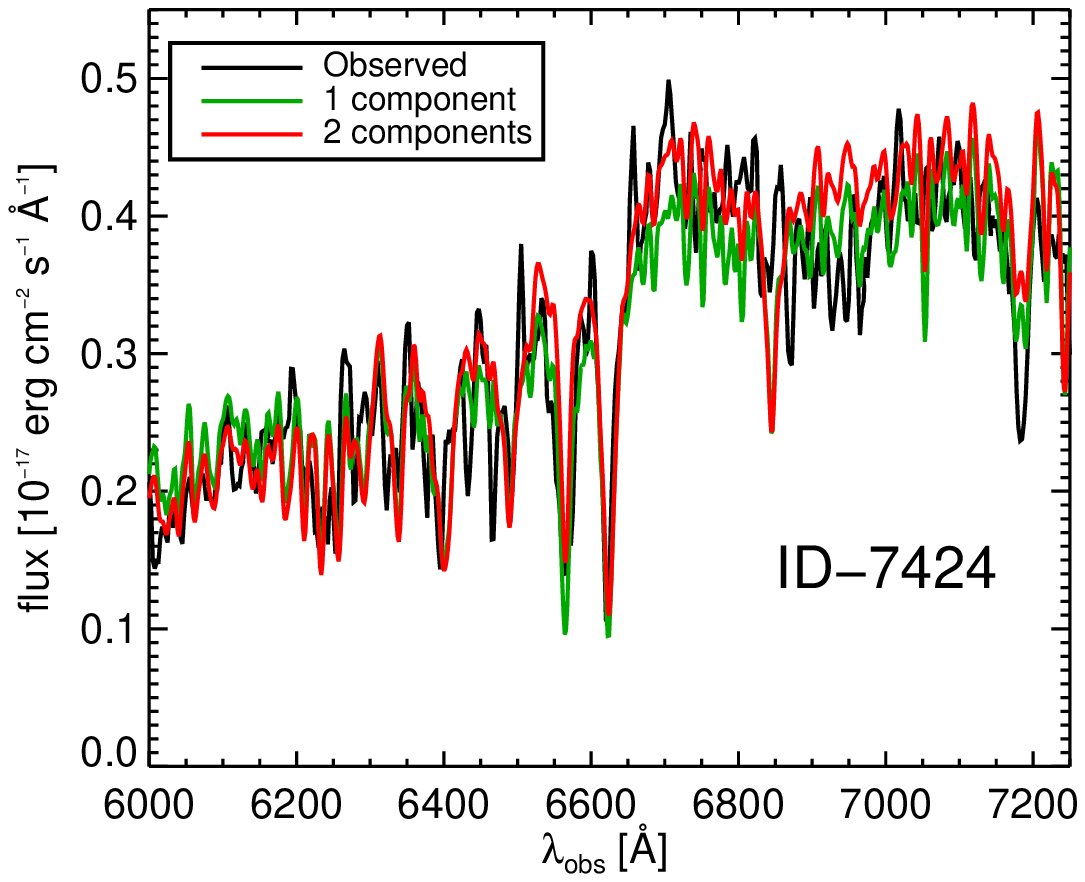}
\caption{Left-hand panels: comparison between single-component solutions (red line) and double-component models (black line) on the observed SED (blue points). Right-hand
panels: comparison between single-component spectrum (green) and double-component spectrum (red) on the observed spectrum (black) in the region of the $4000$\AA $ $
rest frame.
Upper panels: galaxy ID-9066. Analysis: the corresponding stellar age of this object from the SED fitting analysis was $2$ Gyr (with $\tau=0.3$ Gyr and
A$_v=0.6$, see Table \ref{tab:fotom}); instead, from the index measurements we obtained an age estimate of about $0.9$ Gyr from the $\Delta4000$ index and none from the
H+K(Ca{\small\texttt{II}}) index because its value is above the critical value (i.e. $1.2$). We thus proceeded with the double-component analysis on models with gas emission lines. The result
is a composite stellar population with a $6$ Myr younger component (with signs of recent star formation) reddened with A$_v=2.8$, superimposed on the bulk of the stellar
mass ($>99.3$ per cent) with an older age of $1.7$ Gyr. The value of the older component age is similar to the photometric age, but the single-component solution
was not consistent with the measured indices. Instead the double-component solution, with the adding of small amounts of very young stellar population
has been able to adjust the fit of the spectral features, and taking into account at the same time the trend of the whole SED.
Bottom panels: galaxy ID-7424. Analysis: from the SED fitting analysis this object was described by a $2.6$ Gyr old stellar population with $\tau=0.4$ Gyr and
A$_v=0.2$, on the other hand, H+K(Ca{\small\texttt{II}}) and $\Delta4000$ indices pointed to ages of $0.7$ and $1.9$ Gyr respectively (see Table \ref{tab:ages}). Moreover, the
single-component solution was not consistent with the measure of the H+K(Ca{\small\texttt{II}}) index. The double-component analysis found a consistent solution composed by a main
component with old age $6.25$ Gyr and a relative high fraction (about $15$ per cent) of a young component with age $0.7$ Gyr, reddened with A$_v=0.5$.}
\label{esempidouble}
\end{figure*}

In Figure \ref{esempidouble}, we present two examples of the analysis that we performed on two opposite cases which are not consistent with a single component
description: object ID-9066 (top panels) which has a high H+K(Ca{\small\texttt{II}}) value ($>1.2$), and object ID-7424 (bottom panels) which has a low value of the H+K(Ca{\small\texttt{II}})
index. In the caption of Figure \ref{esempidouble}, we report, as example, the detailed discussion of these two representative cases.
Left-hand panels of Figure \ref{esempidouble} show the comparison between the single-component (black) and double-component (red) best-fitting solutions on the
observed SED (blue points): both the presented double-component models fit better the spectral distributions. Focusing on the $4000$\AA $ $ rest frame region
(right-hand panels), it can be seen that the double-component spectrum (red) reproduces better the observed spectrum (black) than the single component model
(green), in particular the features involved in the H+K(Ca{\small\texttt{II}}) index definition (i.e. the H and K absorption lines).

\subsubsection{The ages of the two stellar components}
\label{theages}

\begin{figure*}
\begin{centering}
 \includegraphics[width=14cm]{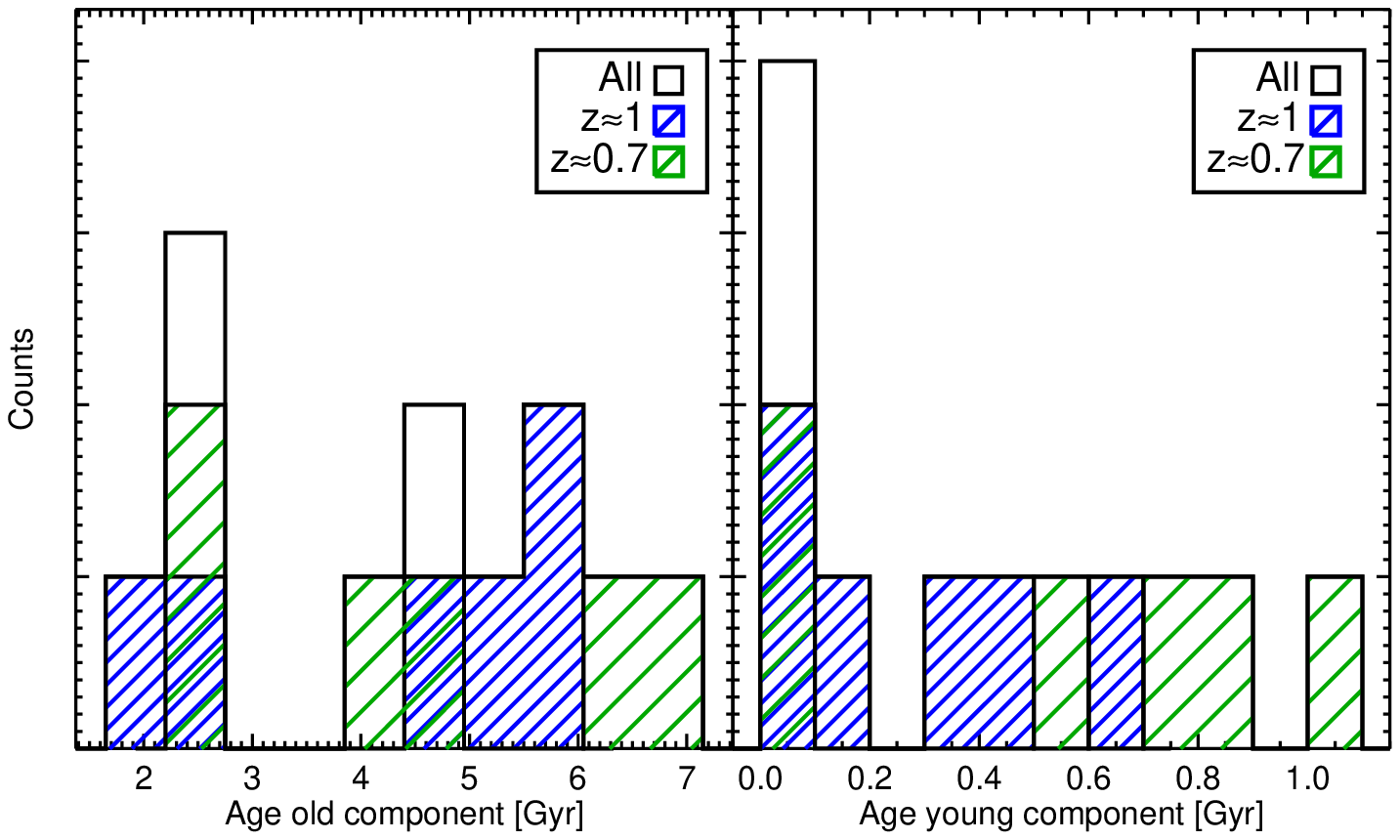}
\caption{Histograms of the double-component analysis results on the $80$ per cent of the sample (see Section \ref{theages}). White: data from all the $80$ per cent of the sample; blue shaded: data from the subsample of galaxies at $z\sim0.7$; green
shaded: data from the subsample of galaxies at $z\sim1$. Left-hand panel: distribution of the ages of the old components in bins of $0.55$ Gyr. There are two visible subsets
with mean age of about $2.25$ Gyr and $5.3$ Gyr. Right-hand panel: distribution of the ages of the young components in bins of $0.1$ Gyr. The mean age is about $0.4$ Gyr.}
\label{istogrammi}
\end{centering}
\end{figure*}

\begin{figure}
\includegraphics[width=9cm]{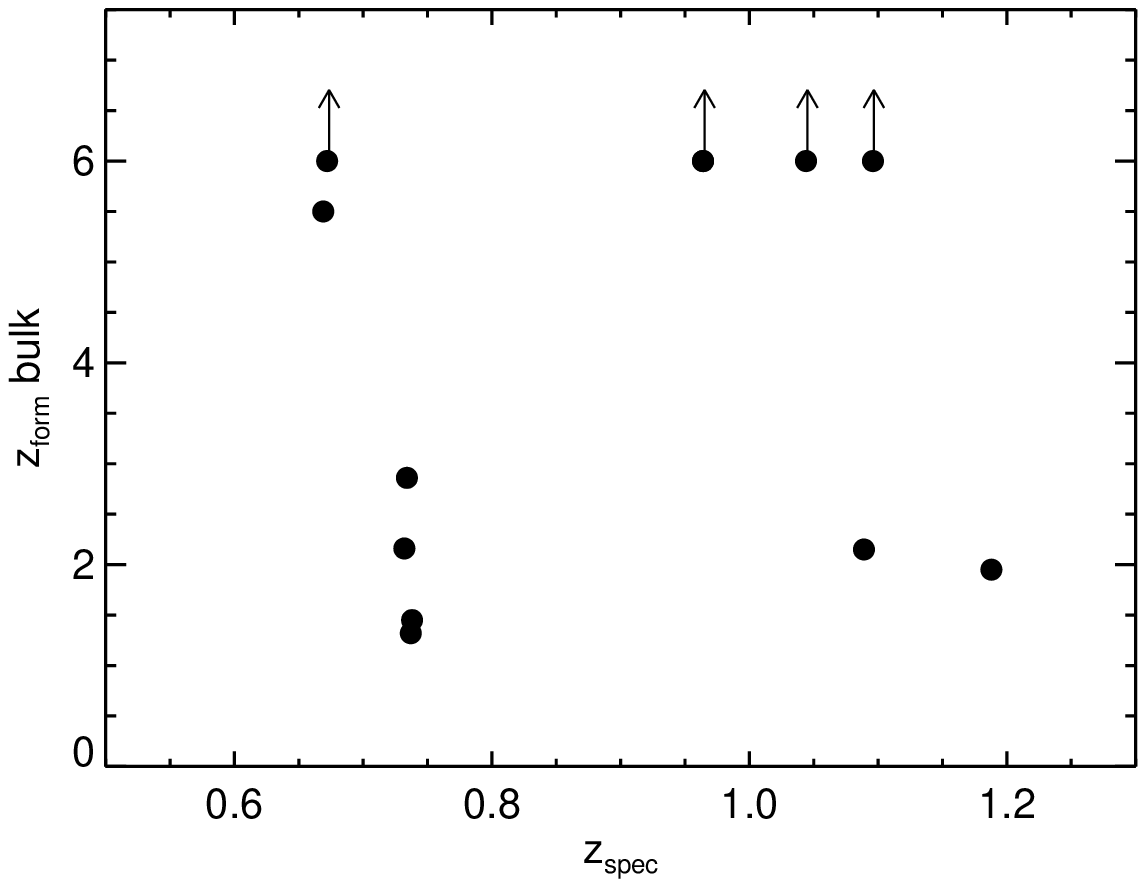}
\caption{$z_{form}$ of the main component as a function of the $z_{spec}$ of the $80$ per cent of the sample of galaxies. Arrows mean that the values can be higher than what indicated by the
respective points.}
\label{zform}
\end{figure}

As we have seen from the analysis of our sample, based on both the spectral indices and the photometric SED fitting, we found evidence that at least $60$ per cent
of the selected ETGs have the bulk of the mass composed of rather old stars and a small percentage of younger stars. In the following, we include also the three
objects (IDs: 1382, 11539 and 9792) for which we have corrected the age of the bulk of stars by means of the index values; indeed, the amounts of younger component, although
not necessary, are found to be very small. In this $80$ per cent of our sample, which we remark here is not complete, with the exception of object ID-7424, we find that the mass fractions of the younger component are small,
less than $10$ per cent, with the smallest being $0.05$ per cent, while the bulk of the mass belongs to the older component. The distributions of the ages of the old
(left-hand panel) and young (right-hand panel) components of this part of the sample, are shown in Figure \ref{istogrammi}. White data represent the whole $80$ per cent of the sample,
while blue and green shaded data represent subsamples at respectively $z\sim0.7$ and $z\sim1$.  Considering the old component age distribution (left-hand panel), it appears bimodal with a younger sample with a mean age
of 2.25 Gyr and with an older one of 5.3 Gyr. This means that the bulk of their stars formed at $1<z_{form}<3$ and $z>5$, as it can be seen in Figure
\ref{zform} which clearly shows that both the two subsamples of galaxies at $z_{spec}\sim0.7$ and $z_{spec}\sim1$ have a $z_{form}$ of their main stellar component bimodally
distributed, without any dependence on redshift. We remark that this result is evidence coming from a casually selected sample of ETGs and hence it cannot be
interpreted as a general bimodality in the distribution of the star formation episodes in ETGs.

On the contrary, in the young component histogram (Figure \ref{istogrammi}, right-hand panel), the distribution is homogeneous with a mean age of $0.4$ Gyr. From both
histograms, it is clear that there is no dependence on redshift in the obtained results, so we cannot identify a single confined cosmic period for the formation
of the stars neither of the old nor of the younger components.

\begin{figure}
\includegraphics[width=9cm]{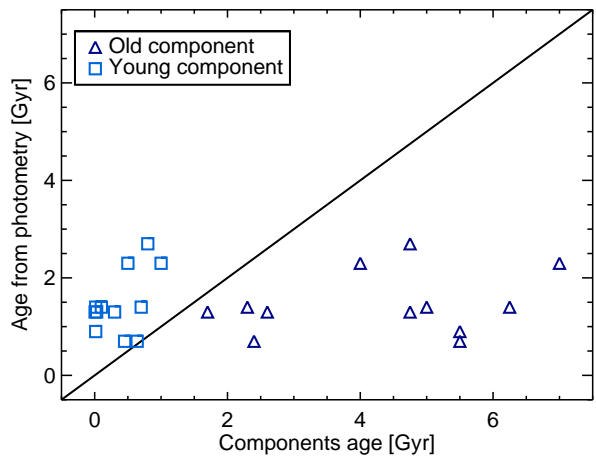}
\caption{Comparison between the ages derived from the SED fitting photometric analysis (from Table \ref{tab:ages}) and the ages of the main component (dark blue triangles) and the young component
(light blue squares) extracted from the double-component analysis (from Table \ref{tab:results}). Data concern the $80$ per cent of the sample.}
\label{confrontorisultati}
\end{figure}

It is worthy to note that the ages of the bulk components are generally older than the ages derived from the photometric analysis.
Indeed, the main components generally reveal the same stellar ages predicted from the values of the $\Delta4000$ index, which, as already discussed, is a measure of the global
stellar population (see Table \ref{tab:results} in comparison with Table \ref{tab:ages}).
In Figure \ref{confrontorisultati} we report the comparison between the ages estimated from the SED analysis (with fixed $\tau=0.1$ Gyr) with
respect to both those of the young components (light blue squares) and those of the old components (dark blue triangles). It can be seen that ages derived from the SED analysis are all
younger than $2.7$ Gyr, while in our double-component analysis, the ages of the old components reach $7$ Gyr. This figure demonstrates that the photometric analysis, without the
constraints imposed by the spectrophotometric indices, clearly underestimates the age of the bulk of stars in ETGs at $z\sim1$. As a consequence, it results that also the
stellar mass derived from the SED fitting is on average underestimated with respect to the value that can be derived from the double-component analysis. Indeed, we
estimate that the mean correction factor for the stellar mass of our sample galaxies is on average $1.2$ with peaks up to a factor of $2$, and the more is the difference between
the age extracted from the SED fitting and the age of the older component, the higher is the value of this correction.

In conclusion, our results suggest that the bulk of the star formation in our sample ETGs at $z\sim1$ happened in two indicative cosmic periods: one at $z_{form}>5$ and the
other more recent at $1<z_{form}<3$. New and minor events of star formation occurred in these galaxies within $1$ Gyr (upper limit) from their observation, i.e. $z\sim1$ and $z\sim0.7$.
Again, we point out that these results come from a limited sample without any statistics and completeness in $z$, so they should be consolidated by means of analysis on larger
samples.

It is worthy to note that the obtained results are strictly related to the peculiarities of the two spectrophotometric measured indices. In particular, the H+K(Ca{\small\texttt{II}})
index has been of fundamental importance to detect the presence of small fractions of young stellar populations also when the bulk of the mass ($>98-99$ per cent) belonged to
the older main component.
Indeed, as it appears evident from Figure \ref{confrontodouble}, the H+K(Ca{\small\texttt{II}}) index experiences large variations even when only a small percentage of stellar mass
younger than $1$ Gyr is added to the old bulk of stars.

\section{Discussion}
\label{discussione}

In the last years, only two works based on high-redshift ETG optical spectra reported a similar spectroscopic analysis based on spectrophotometric indices.
In the following, we try to compare our results with the finding of these two works.

The first one is a work by \citet{onodera} where the authors present a spectral analysis of a sample of $18$ passive elliptical galaxies at $1.4<z<1.8$ in the \emph{Cosmological Evolution Survey} field. In
particular, based on near-IR \emph{Subaru/Multi-Object Infrared Camera and Spectrograph} spectra (resolution R $\simeq500$), they performed the measurements of two spectrophotometric indices in the same
region of our analysis: the H$\delta$ index (defined as H$\delta_F$ by \citealt{wortheyott}) and D$_{n}4000$, a $4000$\AA $ $ break index definition by \citet{balogh},
not coincident with the one adopted in our present work.

To compare our sample with that of \citet{onodera}, we have measured the values of the H$\delta_F$ and D$_{n}4000$ indices on the spectra of our sample.
Results are listed in Table \ref{tab:indicionodera}.

\begin{table}
 \centering
 \caption{Values of the D$_{n}$4000 and H$\delta_F$ indices (used in \citealt{onodera}) and H$\delta_A$ index (used in \citealt{jorgensen}) measured on our sample spectra.}
 \begin{tabular}{lcccc}
 \hline
 ID-MUSIC & $z$ & D$_{n}$4000   & H$\delta_F$   & H$\delta_A$\\
 \hline
 1192 & 1.089 & 1.59$\pm$0.03 & 2.44$\pm$0.53 &  4.88$\pm$0.19 \\
 1382 & 0.964 & 2.10$\pm$0.06 & 2.15$\pm$0.30 &  0.21$\pm$0.60 \\
 1950 & 1.044 & 1.53$\pm$0.08 & 2.16$\pm$0.85 &  0.48$\pm$0.23 \\
 1837 & 0.964 & 2.28$\pm$0.04 & 1.54$\pm$0.34 & -1.85$\pm$0.24 \\
 2694 & 1.135 & 1.96$\pm$0.07 & 1.57$\pm$0.79 &  1.13$\pm$0.53 \\
 9066 & 1.188 & 1.71$\pm$0.15 & 4.12$\pm$0.62 &  2.78$\pm$3.19 \\
 11539& 1.096 & 1.91$\pm$0.06 & 0.05$\pm$1.18 & -0.70$\pm$1.20 \\
 10020& 0.738 & 1.78$\pm$0.09 & 1.90$\pm$1.25 & -0.14$\pm$1.95 \\
 10960& 0.737 & 1.72$\pm$0.05 & 0.15$\pm$0.56 & -1.49$\pm$0.83 \\
 11225& 0.736 & 1.64$\pm$0.04 & 2.06$\pm$0.75 &  1.29$\pm$1.13 \\
 9792 & 0.734 & 1.87$\pm$0.05 & 0.94$\pm$0.69 & -0.85$\pm$1.01 \\
 13386& 0.734 & 1.61$\pm$0.04 & 1.58$\pm$0.85 &  0.05$\pm$1.30 \\
 9838 & 0.732 & 1.80$\pm$0.08 & 1.54$\pm$1.42 & -0.90$\pm$2.20 \\
 17044& 0.672 & 1.93$\pm$0.03 & 1.09$\pm$0.38 &  1.70$\pm$0.54 \\
 7424 & 0.669 & 1.75$\pm$0.09 & 0.62$\pm$1.52 &  1.20$\pm$2.15 \\
 \hline
\end{tabular}
\label{tab:indicionodera}
\end{table}

\begin{figure*}
\includegraphics[width=8.6cm]{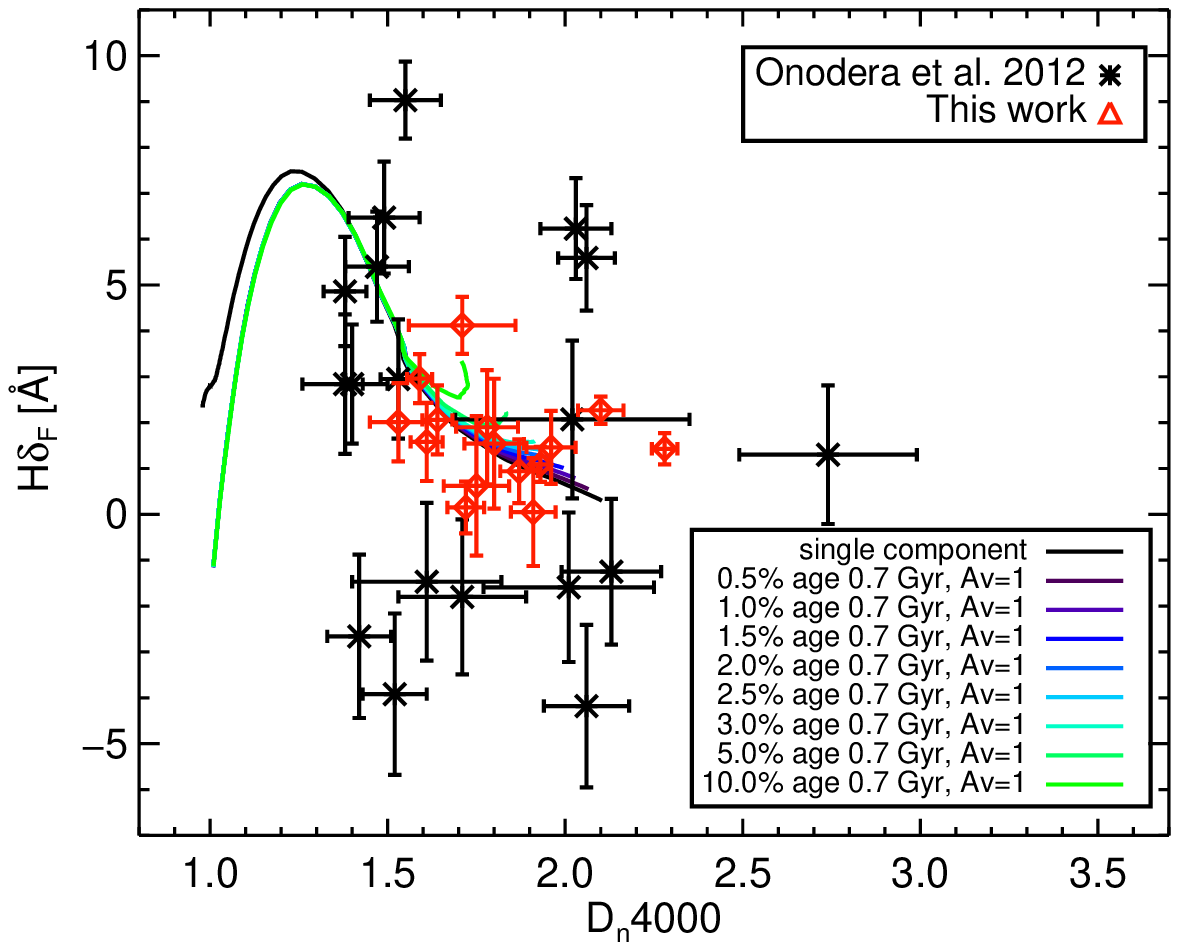}
\includegraphics[width=8.6cm]{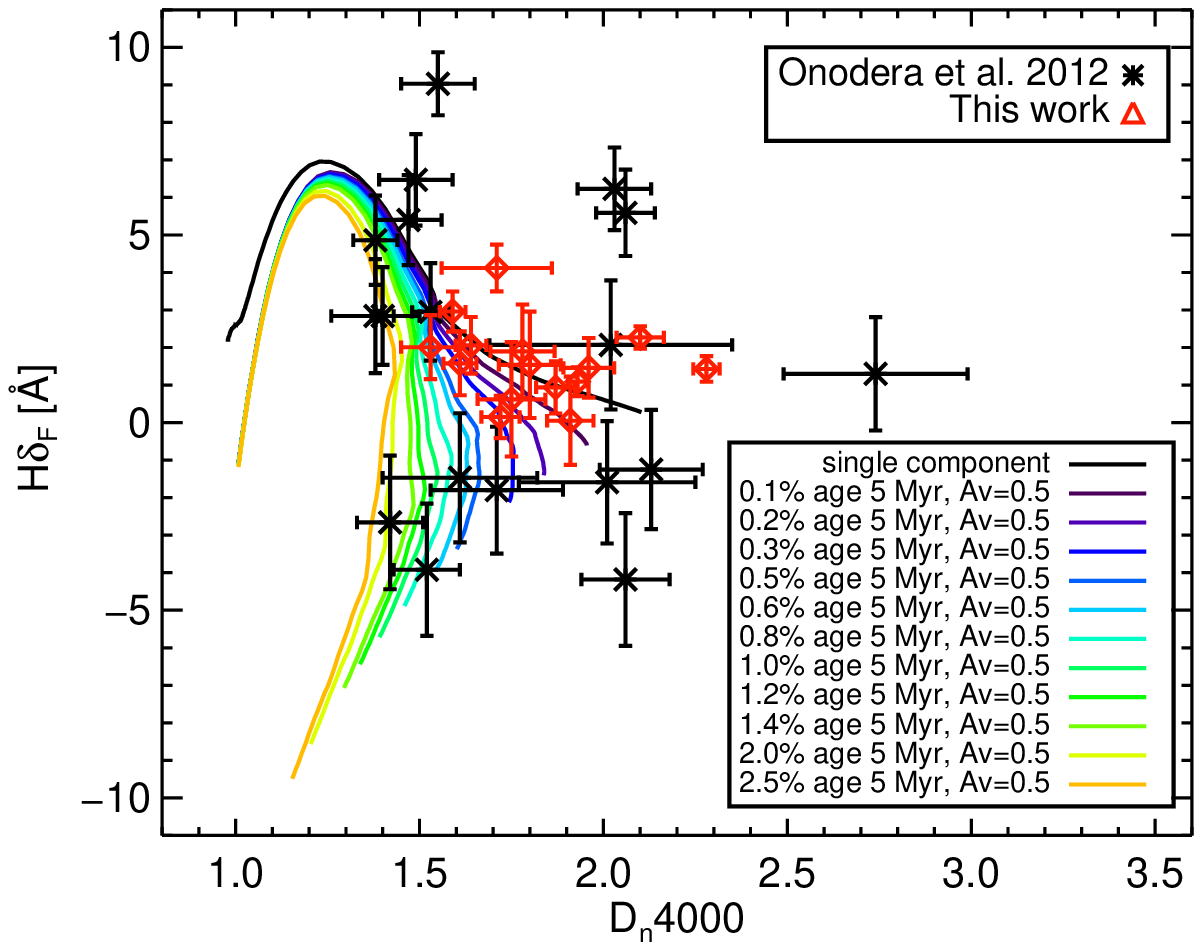}
\includegraphics[width=8.6cm]{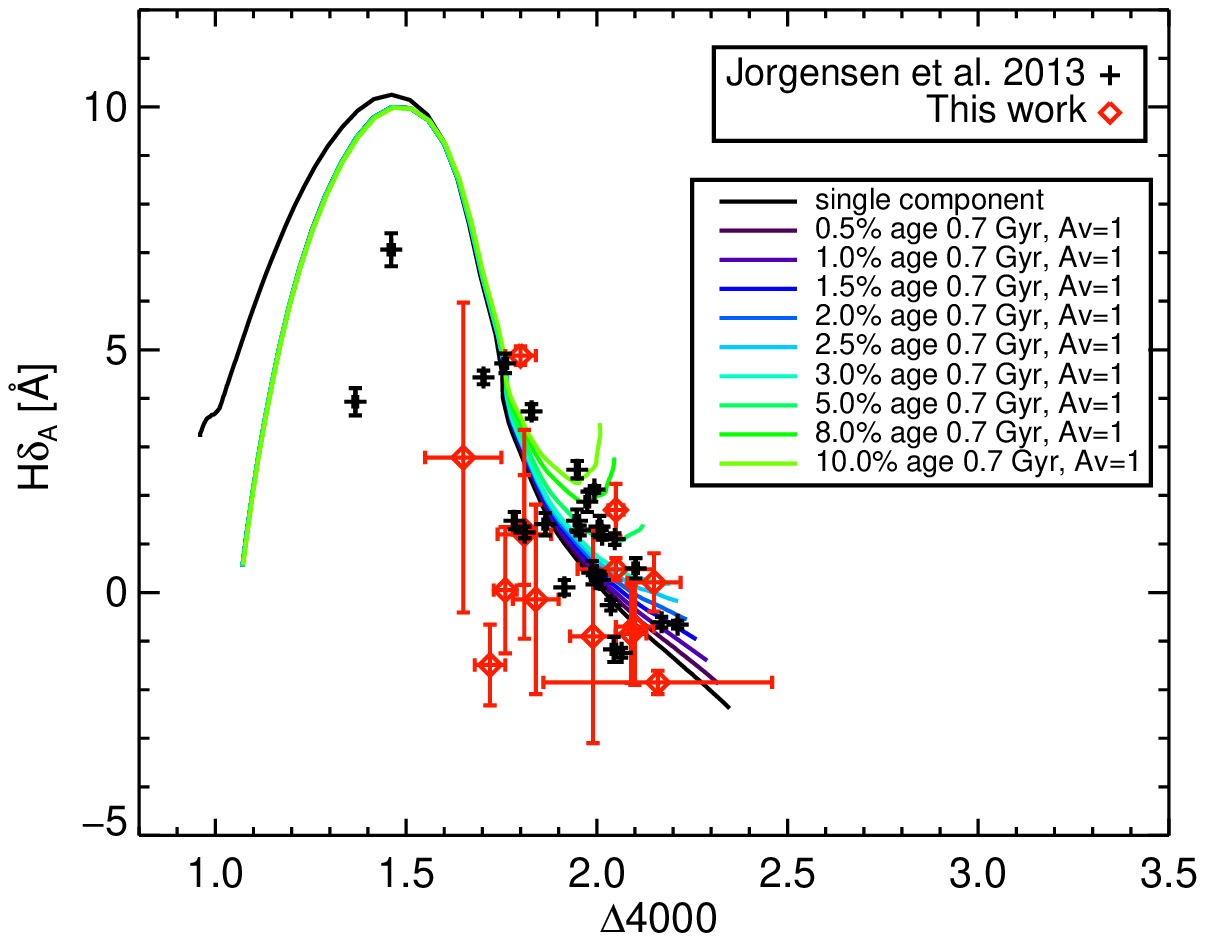}
\includegraphics[width=8.6cm]{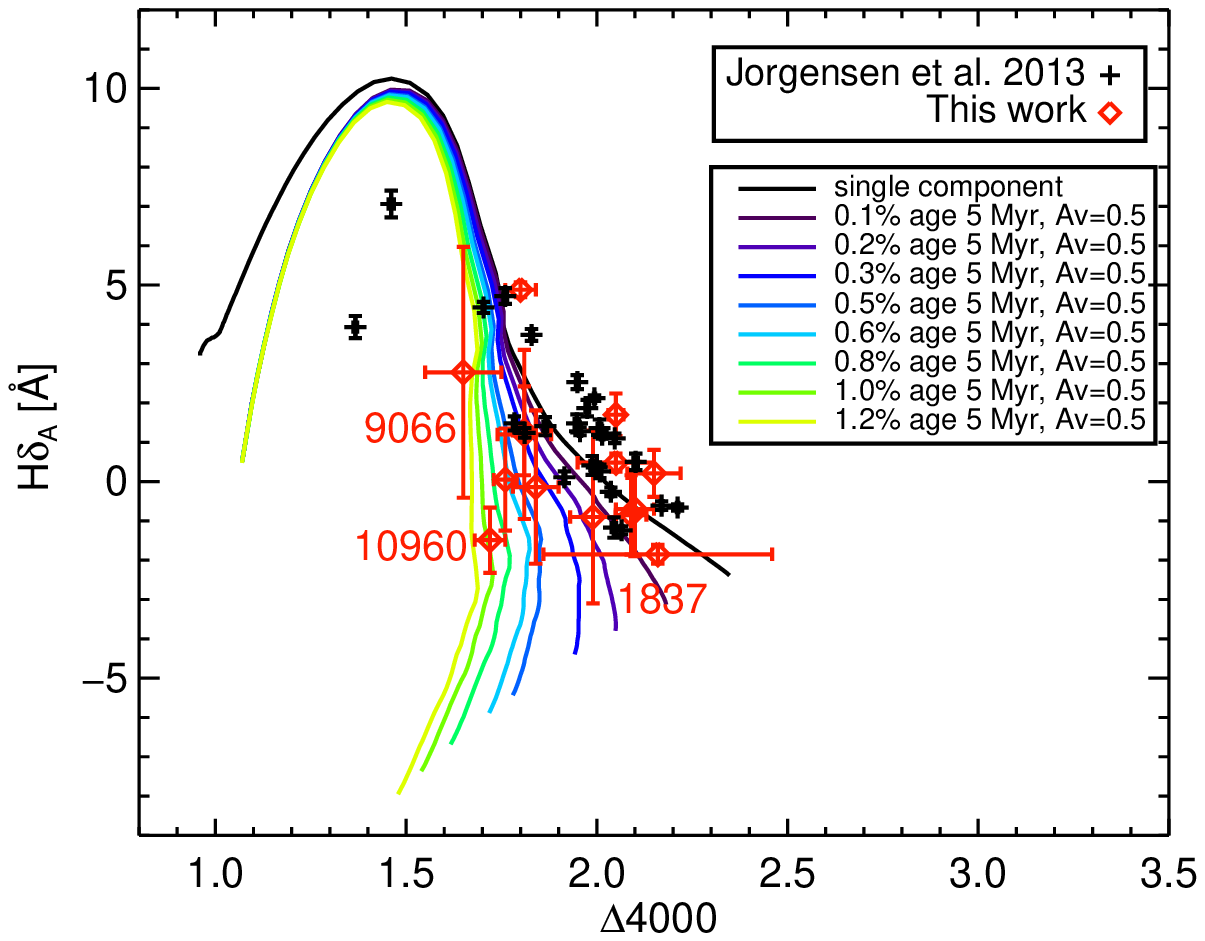}
\caption{Same plots as Figure \ref{confrontodouble} in the H$\delta_F$ versus D$_{n}4000$ plane, for both the \citet{onodera} sample (black points) and this work sample
 (red points)(top panels), and in the H$\delta_A$ versus $\Delta4000$ plane for both the \citet{jorgensen} sample (black points) and this work sample
 (red points)(bottom panels). Double-component models (coloured lines) are composed of the bulk of the mass whose ages increase from left to right (from about
 $10^8$ to $10^{10}$ yr), and by increasing mass fractions of young component with age $0.7$ Gyr with A$_v=1$ (left-hand panels), and with age $5$ Myr, reddened
 with A$_v=0.5$ (right-hand panels). Double-component models in the right-hand panels are shown with the adding of gas emission lines. In all panels, H$\delta_F$ values of
 synthetic models (coloured lines) and data points have been properly corrected to take care of their different spectral resolutions.}
\label{confrontodoubleonoderajor}
\end{figure*}

In their Figure $17$, Onodera and collaborators showed the comparison of their measurements with the predictions of synthetic models (Charlot $\&$ Bruzual, in preparation)
assuming four different star formation histories and most of their objects are not consistent with any of the displayed models. We add our lower redshift sample in the same
H$\delta_F$ versus D$_n4000$ plane, and the result is shown in Figure \ref{confrontodoubleonoderajor} (top panels). Indeed, on average, at fixed D$_{n}4000$ index value, the H$\delta_F$
index value is either higher than expected from models or too low to find any correspondence with models (with $\tau=0.1$ Gyr and solar metallicity, black line in Figure
\ref{confrontodoubleonoderajor}, top panels). We then proceeded with the same arguments followed in the previous sections, considering double stellar components models.
The top-left panel of Figure \ref{confrontodoubleonoderajor} presents composite models with the younger component at fixed age $0.7$ Gyr,
while in the top-right panel the minor component is so young, $5$ Myr, to present emission lines and to produce negative values of the H$\delta_F$ index.
Double-component models with the adding of gas emission are successful in explaining many of the data points of both samples, in particular those with lower values of
both indices. Unfortunately, those points with extreme high values of H$\delta_F$ (and relative high values of D$_{n}4000$ index) remain still unexplained with any
combinations of stellar component ages, as already noticed by \citet{onodera}.

Focusing only on our sample data (red points in Figure \ref{confrontodoubleonoderajor}, top panels), it is worth to stress that the obtained spread in the H$\delta_F$ versus
D$_{n}4000$ index plane is more restrained than that in the H+K(Ca{\small\texttt{II}}) versus $\Delta4000$ index plane discussed in this work (see Figure \ref{confrontomodelli}).
This is due to the lower sensitivity of the H$\delta_F$ index respect to that of the H+K(Ca{\small\texttt{II}}) index to the presence of very small amounts of young stars. As explained
in Section \ref{indici}, indeed, the measure of the H line depth and its ratio with the K line is influenced almost only by the blended H$\epsilon$ line depth, making
the values of their ratio (i.e. H+K(Ca{\small\texttt{II}}) index) a very sensible tools for detecting young stellar populations.
The larger effectiveness of the H+K(Ca{\small\texttt{II}}) index with respect to the H$\delta_F$ index in finding the younger component is easily verified looking at the synthetic models. Colored
lines in Figure \ref{confrontodoubleonoderajor} (top left panel) indicate the values of the H$\delta_F$ and D$_{n}4000$ indices of double-component models with increasing
fractions of the younger component, and it can be noticed that the detachment from the single-component model (black line) is limited with respect to what obtained in Figure
\ref{confrontodouble} in the case of H+K(Ca{\small\texttt{II}}) index.

\begin{figure}
\includegraphics[width=9cm]{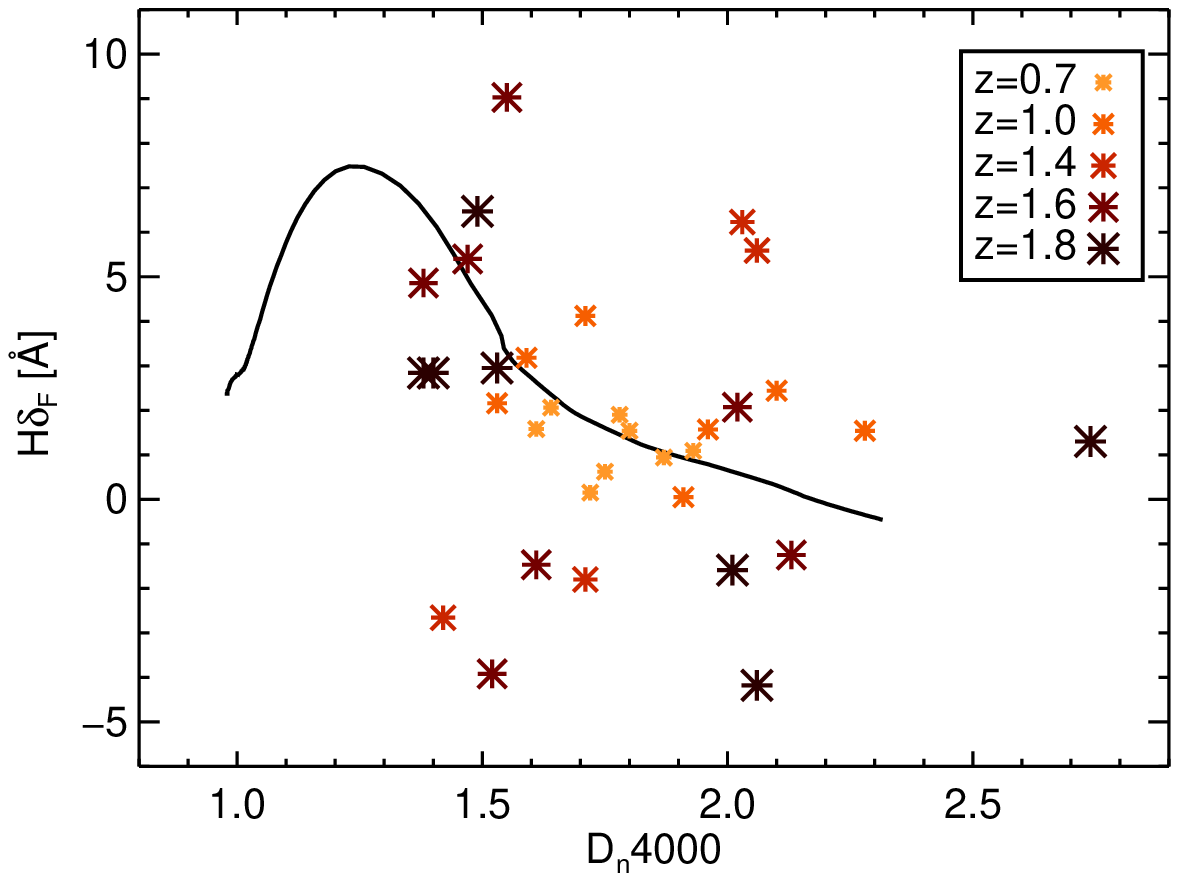}
\caption{H$\delta_F$ versus D$_{n}4000$ plot. Black line is the single-component model with $\tau=0.1$ Gyr and solar metallicity. Data from the sample of \citet{onodera}
and this work galaxies. Data points are divided into five
redshift bins (from yellow to dark red crosses): $z\sim0.7$, $\sim1$, $\sim1.4$, $\sim1.6$ and $\sim1.8$. Errors bars are omitted for simplicity; they would be the
same as in Figure \ref{confrontodoubleonoderajor}.}
\label{variazionez}
\end{figure}

Interestingly, from both top panels of Figure \ref{confrontodoubleonoderajor}, it can be noticed that data points of our sample (red points) stay closer to models than those of
\citet{onodera}. Since our points refer to lower $z$ galaxies, we thus looked for a correlation of this spreading with redshift. We separated the entire sample of galaxies in five subsets with different
range of redshift: $z\sim0.7$, $\sim1$, $\sim1.4$, $\sim1.6$ and $\sim1.8$. In Figure \ref{variazionez}, we present the same plot of Figure \ref{confrontodoubleonoderajor},
but with only a representative single-component model (black line, with $\tau=0.1$ Gyr and solar metallicity). Colours of the data points go from yellow, low-redshift
objects ($z\sim0.7$), to dark red, the highest redshift ($z\sim1.8$). From this plot, it is clear that low-$z$ data stay closer to the model than those with high $z$ and that
the more the redshift increases, the more the data are widespread in the index index plane, i.e. more distant from models. Furthermore, in Figure $17$ in \citet{onodera}, the
authors present the comparison of their high-$z$ data with local SDSS passive galaxies, and the trend seems to be confirmed: local objects are generally well
represented by single-component models with respect to high-$z$ data.
In high-redshift objects, differences of ages among possible multiple stellar components are easier to be detected thanks to the younger mean ages, for which even small age
differences produce strong signatures (e.g. Balmer lines depth), while in the local Universe the same age differences are almost impossible to be detected for the stellar
population ageing, thus vanishing the opportunity of observing the presence of double components. Indeed, two stellar populations with $8$ and $12$ Gyr, respectively, have almost
the same Balmer lines depth. This strengthens the advantage of dealing with direct spectral measurements of higher redshift ETGs in order to study their star formation history.

The second interesting comparison is with the work by \citet{jorgensen}. Their analysis is based on high-S/N optical spectroscopy of the early-type members of three galaxy
clusters at $z>0.5$ in the framework of the ``The Gemini/HST Galaxy Cluster Project'' \citep{jorgensen05}. We limited our comparison to the highest $z$ clusters which are
the RXJ0152.7-1357 at $z=0.83$ and RXJ1226.9+3332 at $z= 0.89$, because they match the average redshift of our sample. For these galaxy clusters, in particular, the authors
make available the measurements of some spectrophotometric indices obtained from the GMOS-N instrument (R $=1918$).
They adopt the same definition of the $\Delta4000$ index as we adopted, and the definition of the H$\delta_A$ by \citet{wortheyott}, different from that of \citet{onodera}.
In order to perform the comparison, we thus measured the values of the H$\delta_A$ of our sample galaxies, and the results are listed in column $5$ of Table \ref{tab:indicionodera}.

Once again we compared the results with the expectation of the single-component BC03 models (black line, with $\tau=0.1$ Gyr and solar metallicity), in Figure
\ref{confrontodoubleonoderajor}, bottom panels. Data points from \citet{jorgensen} (black points) remain relative close to the single-component model predictions
(black line), in agreement with the measure of our sample, supporting the previous finding that the secondary younger component is more evident at increasing $z$.
In order to preserve the graphic clearness, in Figure \ref{confrontodoubleonoderajor} only data points belonging to RXJ1226.9+3332 galaxy cluster are shown.
In Figure \ref{confrontodoubleonoderajor}, bottom-left panel, we report the comparison of both the two sample data with the double-component models (colored lines), the
same as in the top-left panel: points with higher values of H$\delta_A$ index at fixed $\Delta4000$ index value result fully explained by adding a small mass fraction of
young stars. On the other hand, only some of our sample data points (red points) have lower values of H$\delta_A$ index at fixed $\Delta4000$ index and
they can be again explained with double-component models adding very young stellar components displaying Balmer emission lines. In particular, it is interesting to note
that the three objects ID-9066, ID-10960 and ID-1837 of our sample, highlighted with labels in Figure \ref{confrontodoubleonoderajor} (bottom-right panel), have H$\delta_A$ and
$\Delta4000$ indices which are consistent with double-component models displaying emission lines in agreement with the results found using the H+K(Ca{\small\texttt{II}}) index with respect to
the H$\delta_A$ one in this work (see Figure \ref{confrontodouble}, right-hand panel).

Finally, we remark that in the bottom panels of Figure \ref{confrontodoubleonoderajor}, we are comparing in the same plots data coming from two samples of galaxies at about the
same mean redshift but in different environments: our sample (red points) is composed of field galaxies, i.e. low-density environment (with mean $z=0.83$), while
\citet{jorgensen} sample (black points) is composed exclusively of galaxies belonging to the cluster RXJ1226.9+3332 at $z= 0.89$. The fact that the galaxies of the sample of
\citet{jorgensen} do not need on average the presence of a second stellar component modelled with Balmer emission lines, means that in general their stellar populations are 
homogeneous, i.e. formed in a single or very short burst with no secondary bursts,
in agreement with the well-known results of \citet{thomas}, where high-density environment galaxies are expected to form their star earlier than low-density counterparts
and in a shorter time-scale (see Figure 10 in \citealt{thomas}). Supposedly, many cluster objects have experienced the same mechanism of star formation and assembling of our sample
ETGs that led to the presence of stellar components with different properties, but these events must have occurred earlier in the cosmic time and are less appreciable at the
observation redshift.

\section{Summary and conclusions}
\label{conclusione}
We presented a spectroscopic analysis based on the measurements of age-dependent specrophotometric indices in the region of $4000$\AA $ $ rest frame on high-redshift ETGs.
We based our analysis on all the up-to-now publicly available optical spectra of ETGs at $z\sim1$ in the GOODS-South field, with ascertained morphological classification (see Tamburri et al., in
preparation), and with sufficient high-S/N ratio in the region of $4000$\AA $ $ break rest frame to allow a detailed spectral analysis. The
sample was then composed of $15$ ETGs at $0.7<z<1.1$, whose spectral data came from three observing campaigns performed by \citet{vdwel}, \citet{mignoli} and
\citet{popesso}.

From the reduced monodimensional spectra, we have measured the values of the two age-dependent spectrophotometric indices: H+K(Ca{\small\texttt{II}}) and $\Delta4000$. The choice of these indices was
supported by their sensitivity to the presence of multiple stellar populations with different age. In particular, we have stressed the peculiarity of the H+K(Ca{\small\texttt{II}}) index of being
very sensitive to age variations even in the case of small amounts of very young stellar content.

From the comparison of the index values with the expectations of single-component BC03 models, we found that many measured values deviate significantly from models.
Nor synthetic models with different star formation time-scale $\tau$ neither those with different metallicity could explain the discrepancy from the ages indicated by the
$\Delta4000$ index (in general older) and those pointed out by the H+K(Ca{\small\texttt{II}}) index. Furthermore, particularly intriguing it was the finding of some values of the H+K(Ca{\small\texttt{II}}) index
which were too high to find any correspondence with models.

Such a behaviour has thus been attributed to inhomogeneous properties of the stellar populations in the sample galaxies, in particular to age variations. Simple models
where the bulk of the star formation happened in a dominant initial event (within a time-scale $\tau$) are not able to explain our
results: a further degree of complexity is necessary to understand the star formation history of our sample galaxies.
To take into account these inhomogeneities, we then introduced in the modelling a second star formation event, thus considering the stellar population composed of two stellar
components which differ from their ages. Adopting composite models, the result is that a small mass percentage of younger stellar component is enough to explain those
index values that were not consistent with single-component models. In particular, those high values of the H+K(Ca{\small\texttt{II}}) index could be explained introducing the contribution
of gas emission lines, due to the presence of star forming regions. Indeed, small mass fractions of very young components ($<0.1$ Gyr), where the star formation is
still ongoing and the H$\epsilon$ Balmer line is in emission, are able to fill the whole H line involved in the H+K(Ca{\small\texttt{II}}) index, thus increasing its expected value.
From the fitting analysis we obtained that for at least $60$ per cent of the galaxies of our sample there are pieces of evidence of the presence of small mass fractions of a younger population coexisting with
a much older stellar bulk.

Data suggest that the ages of the \emph{older} components follow two distinct distributions with mean ages of $2.25$ and $5.3$ Gyr without any correlation with $z$, which correspond to mean formation
redshift for the bulk of stars at $1<z_{form}<3$ and $z_{form}>5$. However, due to the limited number of galaxies of our sample ($15$) not uniformly distributed in redshift,
these results could be only a selection effect. Moreover, we found that the main component ages tend to be older than what concluded from the standard
SED fitting analysis (in agreement with \citealt{fan2014} analysis on star-forming galaxies), with some implications on the stellar mass estimate, which results to be underestimated by a
factor of $\sim1.2$ on average.
The following star-forming events (i.e. the \emph{younger} components) are detected at any redshift of the galaxies of our sample ($0.7<z<1.1$), so it cannot be outlined in a
common cosmic period in which the younger component has been accreted. It appears more likely that the star formation is triggered steadily over the cosmic time within field
ETGs.

We have supposed the presence of younger components also in higher $z$ sample of passive elliptical \citep{onodera} up to $z=1.8$, and in $z\sim1$ cluster galaxies
\citep{jorgensen} even if based on the H$\delta$ index rather than the H+K(Ca{\small\texttt{II}}) index, the former less sensible to small
amounts of young stellar content. We have found that in general, the ages of the minor younger component are less young for galaxies belonging to cluster, and more extreme
with the increasing of redshift.

Having revealed the presence of small young stellar components in many galaxies in a wide redshift range ($0.7<z<1.8$, thanks to the enlarged sample) implies that there
must be a common mechanism by which ETGs at any cosmic time either accrete new amounts of young stellar mass by means of minor merging events, or activate new events of star
formation almost constantly over time.
Minor merger-induced star formation is the most widely accredited hypothesis to explain the presence of ongoing star formation in
ETGs, since it has been also found in the local Universe \citep{kaviraj2012}. Minor mergers are those with mass ratio ($M/m$) in the range $1/10 < M/m < 1/4$, and our
findings on the younger components mass fractions are consistent with this scenario. Indeed, adopting the assumption of ``dry'' minor merging, which in general involves a maximum
value of gas fraction of $10$ per cent of the total accreted mass (stars+gas), the very small fractions of the second components observed in this work, i.e. $\sim0.1$
per cent,
result in agreement with this description. On the other hand, obviously, also wet minor merging events are able to explain the origin of the observed younger component, with
the accretion of external gas being the driver of new star formation activity.
However, minor merging events do not seem to be so frequent in the cosmic epoch from $z\sim1$ to now. Indeed, \citet{lopez2012}
have found that from $z\sim1$ to $z\sim0$ the number of minor merging per red galaxy is $0.46\pm0.06$; thus, it seems rather unlikely that in an albeit small sample of $15$
galaxies, we found that at least $9$ objects (i.e. $60$ per cent) have recently experienced a minor merging event. In addition, it must be considered that all these objects have been
classified as ETGs and their morphology appear all regular; thus, if merger events have happened in the cosmic histories of these galaxies, they must have
taken place at least few Gyr earlier with respect to their observation in order to resettle the elliptical morphology. But this is not in agreement with our findings where the younger
component is always observed with age $< 1$ Gyr.

Moreover, other independent works, such that by \citet{garg12}, not only have demonstrated the inhomogeneities of stars in ETGs at $z>1$, but they have also localized the younger
component in the inner part of the galaxies. This naturally leads to the question of what is the physical mechanism which, after the external mass accretion by means of
hypothetical merging events, is able to bring only the younger stars to the centre of galaxies at least in the case of dry merger event.

On the other hand, our finding of a second younger component also in other high-$z$ samples suggests that the presence of a small percentage of young stars at any epoch which coexist with
the passively evolving older stellar bulk of the galaxy could be common in ETGs. In fact, it is known that ETGs contain small
fraction of cold molecular gas that could be converted in star formation activity and its presence has been observationally detected in the local
Universe \citep{lees91, crocker2011, panuzzo2011}. This minimal gas reservoir, besides having a presumed external origin, could also be due to the stellar mass loss or
may be left over from the initial star formation major burst in the earlier epochs. The favoured scenario is that of the cold accretion: it has been shown
\citep{fardal01,keres05,dekel06} that a cold gas mode could be responsible for the star formation in the cosmic history of galaxies and that it can be present, with its
filamentary nature and in small quantities, also in spheroidal passive systems where the hot gas mode is dominant. In fact,
\citet{keres05} show that, from $z\sim3$ to $z\sim0$ hot gas mode dominated galaxies, i.e. passive and massive spheroids, can form new stars thanks to
small amounts of cold gas, and the involved mass fractions, $<20$ per cent for stellar masses log($M_*$)$>10.5$, are fully consistent with the detected mass fractions of the younger
component in our analysis sample. In addition, this scenario could be in agreement with the observed inner position of the younger component, as the cold
gas stream is assumed to infall towards the centre of galaxy attracted by the main potential well.

Thus, the detected young stellar component in the sample of ETGs analysed in the present work suggests that a star formation activity took place steadily over time during
the secular evolution of these galaxies, activated by small gas quantities since $z>1$.

\section*{Acknowledgements}
We thank the anonymous referee of this paper for providing constructive comments that improved the manuscript. This work has received financial support from PRIN-INAF
($1.05.09.01.05$).

\label{lastpage}


\begin{thebibliography}{99}

\bibitem[\protect\citeauthoryear{Balogh et al.}{1999}]{balogh}
Balogh M. L., Morris S. L., Yee H. K. C., Carlberg R. G., Ellingson E., 1999, ApJ, 527, 54
\bibitem[\protect\citeauthoryear{Bruzual}{1983}]{bruzual83}
Bruzual G., 1983, AJ, 273, 105
\bibitem[\protect\citeauthoryear{Bruzual \& Charlot}{2003}]{bc03}
Bruzual G., and Charlot S., 2003, MNRAS, 344, 1000 (BC03)
\bibitem[\protect\citeauthoryear{Calzetti et al.}{2000}]{calzetti}
Calzetti D., Armus L., Bohlin R.C.,  Kinney A.L., Koornneef J., and Storchi-Bergmann T., 2000, ApJ, 533, 682
\bibitem[\protect\citeauthoryear{Cappellari et al.}{2009}]{cappellari2009}
Cappellari M., et al., 2009, ApJ, 704, L34
\bibitem[\protect\citeauthoryear{Chabrier}{2003}]{chabrier}
Chabrier G., 2003, PASP, 115, 763
\bibitem[\protect\citeauthoryear{Charlot $\&$ Longhetti}{2001}]{charlot2001}
Charlot S., and Longhetti M., 2001, MNRAS, 323, 887
\bibitem[\protect\citeauthoryear{Cimatti et al.}{2004}]{cimatti2004}
Cimatti A., et al., 2004, Nature, 430, 184
\bibitem[\protect\citeauthoryear{Coccato et al.}{2010}]{coccato2010}
Coccato L., Gerhard O., and Arnaboldi M., 2010, MNRAS, 407, L26
\bibitem[\protect\citeauthoryear{Crocker et al.}{2011}]{crocker2011}
Crocker A.F., Bureau M., Young L.M., and Combes F., 2011, MNRAS, 410, 1197
\bibitem[\protect\citeauthoryear{Daddi et al.}{2005}]{daddi2005}
Daddi E., et al., 2005, ApJ, 626, 680
\bibitem[\protect\citeauthoryear{Dekel $\&$ Birnboim}{2006}]{dekel06}
Dekel A., and Birnboim Y., 2006, MNRAS, 368, 2
\bibitem[\protect\citeauthoryear{Fan et al.}{2014}]{fan2014}
Fan L., Lapi A., Bressan A., Nonino M., De Zotti G., and Danese L., 2014, RAA, 14, 15
\bibitem[\protect\citeauthoryear{Fardal et al.}{2001}]{fardal01}
Fardal M.A., Kats N., Gardner J.P., Hernquist L., Weinberg D.H., and Dave' R., 2001, ApJ, 562, 605
\bibitem[\protect\citeauthoryear{Gargiulo et al.}{2012}]{garg12}
Gargiulo A., Saracco P., Longhetti M., La Barbera F., and Tamburri S., 2012, MNRAS, 425, 2698
\bibitem[\protect\citeauthoryear{Glazebrook et al.}{2004}]{glazebrook2004}
Glazebrook K., et al., 2004, Nature, 430, 181
\bibitem[\protect\citeauthoryear{Guo et al.}{2011}]{guo2011}
Guo Y., et al., 2011, ApJ, 735, 18
\bibitem[\protect\citeauthoryear{Hamilton}{1985}]{hamilton85}
Hamilton D., 1985, AJ, 297
\bibitem[\protect\citeauthoryear{Huang et al.}{2013}]{huang2013}
Huang S., Ho L.C., Peng C.Y., Li Z., and Barth A.J., 2013, ApJ, 766, 47
\bibitem[\protect\citeauthoryear{J$\o$rgensen et al.}{2005}]{jorgensen05}
J$\o$rgensen I., Bergmann M., Davies R., Barr J., Takamiya M., and Crampton D., 2005, AJ, 129, 1249
\bibitem[\protect\citeauthoryear{J$\o$rgensen \& Chiboucas}{2013}]{jorgensen}
J$\o$rgensen I., and Chiboucas K., 2013, AJ, 145, 77
\bibitem[\protect\citeauthoryear{Kaviraj et al.}{2012}]{kaviraj2012}
Kaviraj S., et al., 2012, MNRAS, 423, 49
\bibitem[\protect\citeauthoryear{Keres et al.}{2005}]{keres05}
Keres D., Kats N., Weinberg D.H., and Dave' R., 2005, MNRAS, 363, 2
\bibitem[\protect\citeauthoryear{Lees et al.}{1991}]{lees91}
Lees J.F., Knapp G.R., Rupen M.P. and Phillips T.G., 1991, ApJ, 379, 177
\bibitem[\protect\citeauthoryear{Longhetti et al.}{1999}]{longhetti99}
Longhetti M., Bressan A., Chiosi C., and Rampazzo R., 1999, A$\&$A, 345, 419
\bibitem[\protect\citeauthoryear{Longhetti \& Saracco}{2009}]{longhetti2009}
Longhetti M., and Saracco P., 2009, MNRAS, 394, 774
\bibitem[\protect\citeauthoryear{Lopez-Sanjuan et al.}{2012}]{lopez2012}
Lopez-Sanjuan C., et al., 2012, A$\&$A, 548, A7
\bibitem[\protect\citeauthoryear{Maraston $\&$ Stromback}{2011}]{ma11}
Maraston C., and Stromback, G., 2011, MNRAS, 418, 2785
\bibitem[\protect\citeauthoryear{Mignoli et al.}{2005}]{mignoli}
Mignoli M. et al., 2005, A$\&$A, 437, 883
\bibitem[\protect\citeauthoryear{Naab et al.}{2009}]{naab2009}
Naab T., Johansson P.H., and Ostriker J.P., 2009, ApJ, 699, L178
\bibitem[\protect\citeauthoryear{Onodera et al.}{2012}]{onodera}
Onodera M., et al., 2012, ApJ, 755, 26
\bibitem[\protect\citeauthoryear{Oser et al.}{2010}]{oser2010}
Oser L., Ostriker J.P., Naab T., Johansson P.H., and Burkert A., 2010, ApJ, 725, 2312
\bibitem[\protect\citeauthoryear{Oser et al.}{2012}]{oser2012}
Oser L., Ostriker J.P., Naab T., and Johansson P.H., 2012, ApJ, 744, 630
\bibitem[\protect\citeauthoryear{Panuzzo et al.}{2011}]{panuzzo2011}
Panuzzo P., Rampazzo R., Bressan A., Vega O., Annibali F., Buson L.M., Clemens M.S., and Zeilinger W.W., 2011, A$\&$A, 528, A28
\bibitem[\protect\citeauthoryear{Popesso et al.}{2009}]{popesso}
Popesso P., et al., 2009, A$\&$A, 494, 443
\bibitem[\protect\citeauthoryear{Renzini}{2006}]{renzini2006}
Renzini A., ARA$\&$A, 44, 141
\bibitem[\protect\citeauthoryear{Rocca-Volmerange et al.}{2013}]{rocca2013}
Rocca-Volmerange B., et al., 2013, MNRAS, 429, 2780
\bibitem[\protect\citeauthoryear{Rose}{1985}]{rose85}
Rose J.A., 1985, AJ, 90
\bibitem[\protect\citeauthoryear{Sanchez et al.}{2011}]{sanchez2011}
Sanchez H.D., et al., 2011, MNRAS, 417, 900
\bibitem[\protect\citeauthoryear{Santini et al.}{2009}]{santini}
Santini P., et al. 2009, A$\&$A, 504, 751
\bibitem[\protect\citeauthoryear{Saracco et al.}{2010}]{34etg}
Saracco P., Longhetti M., and Gargiulo A., 2010, MNRAS, 408, L21
\bibitem[\protect\citeauthoryear{Strazzullo et al.}{2013}]{strazzullo2013}
Strazzullo V., et al., 2013, ApJ, 772, 118
\bibitem[\protect\citeauthoryear{Thomas et al.}{2005}]{thomas}
Thomas D., Maraston C., Bender R., and de Oliveira C.M., 2005, ApJ, 621, 673
\bibitem[\protect\citeauthoryear{Treu et al.}{2002}]{treu2002}
Treu T., Stiavelli M., Casertano S., M$\o$ller P., and Bertin G., 2002, ApJ, 564, L13
\bibitem[\protect\citeauthoryear{Van der Wel et al.}{2005}]{vdwel}
Van der Wel A., Franx M., van Dokkum P.G., Rix H.-W., Illingworth G.D. and Rosati P., 2005, ApJ, 631, 162
\bibitem[\protect\citeauthoryear{Vanzella et al.}{2005}]{vanz05}
Vanzella E., et al., 2005, A$\&$A, 434, 53
\bibitem[\protect\citeauthoryear{Vanzella et al.}{2008}]{vanz08}
Vanzella E., et al., 2008, A$\&$A, 478, 83
\bibitem[\protect\citeauthoryear{Worthey $\&$ Ottaviani}{1997}]{wortheyott}
Worthey G., and Ottaviani D., 1997, ApJS, 111, 377

\end{thebibliography}
\end{document}